\documentclass[aps,prb,twocolumn,groupedaddress,showpacs,eqsecnum,floatfix,nofootinbib]{revtex4-1}
\usepackage{amssymb,amsmath}

\newcommand{\scale}{\ensuremath{\delta\ell}}

\newcommand{\ka}{\ensuremath{\kappa} } 
\newcommand{\vertex}{\ensuremath{\alpha} }
\newcommand{\dvertex}{\ensuremath{\tilde{\alpha} } }

\usepackage{subfigure,graphicx}
\begin{document}

\title{Dynamical stability of the quantum Lifshitz theory in $2+1$ Dimensions}

\author{Benjamin Hsu}
\affiliation{Department of Physics, Princeton University, Princeton, NJ 08544}

\author{Eduardo Fradkin}
\affiliation{Department of Physics, University of Illinois at Urbana-Champaign, 
Illinois 61801-3080, USA}

\date{\today}

\begin{abstract}
The role of magnetic and electric perturbations in the quantum Lifshitz model in $2+1$ dimensions are examined in this paper. 
The quantum Lifshitz model is an effective field theory for quantum multicritical systems, that include generalized 2D quantum dimer models in bipartite lattices and their generalizations. It describes a class of quantum phase transitions between ordered and topological phases in $2+1$ dimensions. Magnetic perturbations break the dimer conservation law.  Electric excitations, whose condensation lead to ordered phases,  have been studied extensively both in the classical 3D  model and in the quantum 2D model. The role of magnetic vortex excitations whose condensation drive these systems into a $\mathbb{Z}_2$ topological phase has been largely  ignored. Recent numerical studies claim that the quantum theory has a peculiar feature: the dynamical exponent $z$ flows continuously and the quantum theory is hence unstable to magnetic vortices. To study the interplay of both excitations, we perform a perturbative renormalization group study to one loop order and study the stability of the theory away from quantum multicriticality. This is done by generalizing the operator product expansion to anisotropic models. It is found that the dynamical exponent does not appear to flow, in contrast to the classical Monte Carlo study. Possible reasons for this difference are discussed at length.
\end{abstract}


\maketitle

\section{Introduction}

In this paper we examine the scaling behavior of the dynamics of quantum dimer models near quantum criticality and their generalizations. Quantum dimer models are 2D strongly correlated systems with a rich phase diagram that includes both ordered and topological phases. The quantum dimer model was introduced by Rokhsar and Kivelson (RK) in 1988\cite{RK1988}  as a model to describe spin liquid states within a large spin gap. Such quantum disordered phases presumably arise in sufficiently frustrated quantum antiferromagnets so that all spin-ordered states are suppressed.\cite{Moessner-2001b,Moessner-2006,Cano-2010,Moessner-2011} Typical examples of frustrated systems that can be described by quantum dimer models include antiferromagnets in transition metal oxides on Kagome and pyrochlore lattices.\cite{Balents-2002,Bergman-2006a,Bergman-2006b,Cabra}

The most interesting feature of the quantum dimer models is the rich variety of phases that are accessible from a quantum multicritical point with special properties. The phase diagrams of these models are known by a combination of strong coupling arguments and from the existence of a special choice of parameters, know as the Rokhsar-Kivelson (RK) point, at which the quantum Hamiltonians take the form of a sum of local projection operators which allows for the determination of their exact ground state wave function (at the RK point only).\cite{RK1988} The resulting ground state wave functions have the form of a liner superposition of dimer configurations with local weights. A consequence of this structure is that the norm of the ground state wave functions of these models (at their RK points)  is equal to the partition function of classical dimer models on the same 2D lattice whose (local) Gibbs weights are the square of the amplitudes of the wave function. For the simplest quantum dimer model, the ground state wave function is equal to the short-range resonating valence bond state.\cite{Anderson-1987,Kivelson-1987}
Because of the close connection between the ground state wave function of generalized quantum dimer models and two dimensional classical statistical mechanical systems,
a great deal of information about the static properties of such models are known. Given the knowledge of the exact ground state wave function, it is also possible to compute the equal-time-correlation functions by mapping them to a 2D classical counterpart. It has also been instrumental in exact computations of the entanglement entropy in two dimensions,\cite{Fradkin2006,Hsu2009,Pasquier2009,Oshikawa2010,Hsu-2010} thus extending the known results for one-dimensional systems.\cite{Callan-1994,Holzhey-1994,Vidal-2003,Calabrese-2004}

 \begin{figure}[hbt]
	\centering
	\subfigure[]{\label{6v} \includegraphics[width=.47\textwidth]{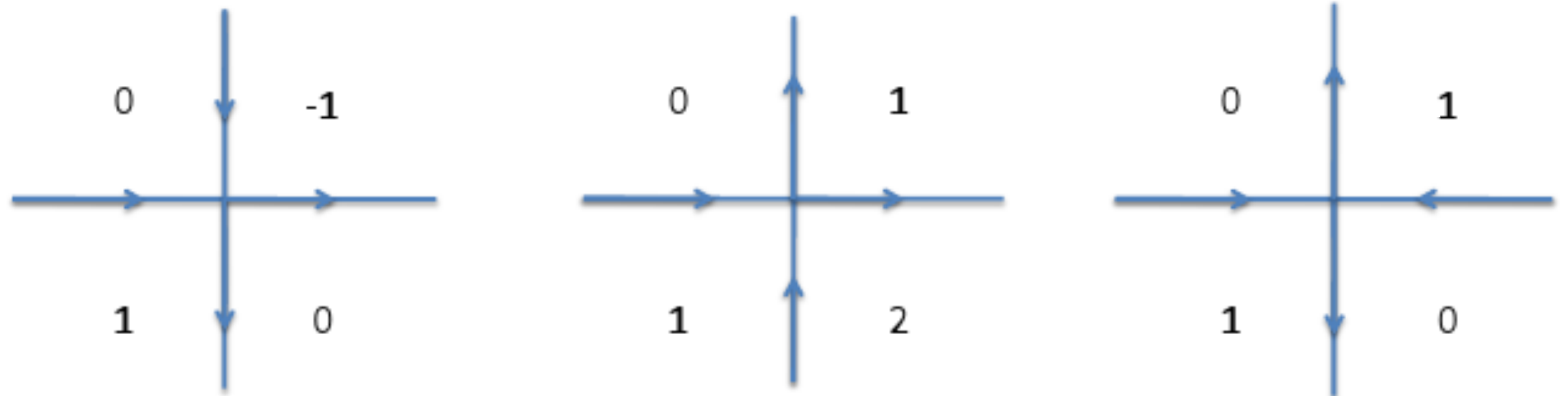} }
	\subfigure[]{\label{8v} \includegraphics[width=.13\textwidth]{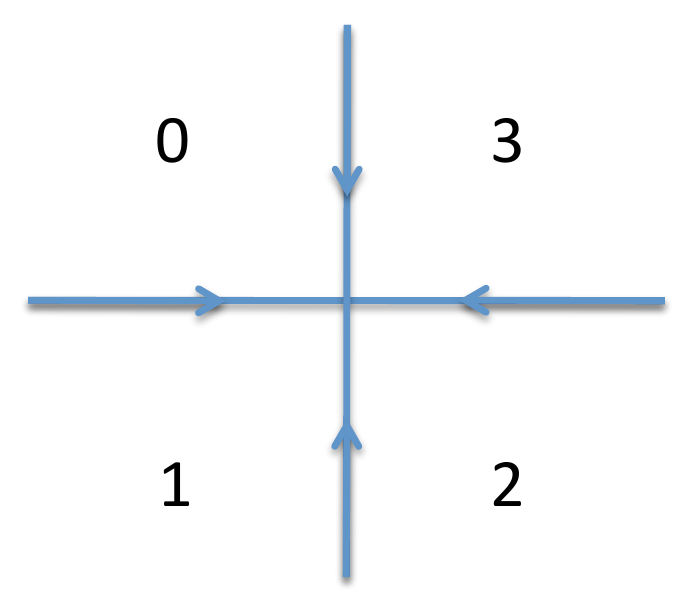} }
	\caption{The amplitude of the wave function depends on the type of vertices that enter in each configuration. Vertices in the (a) six-vertex model with amplitudes (``fugacities'')  $a$, $b$, and $c$, respectively, and (b) the additional vertex in the eight-vertex model with fugacity $d$. These vertices and their inverses with all arrows flipped, form the six vertex and eight vertex models respectively. The heights are displayed in quadrant. To do so, one arbitrarily picks a starting plaquette and moving in a counter clockwise direction, raises the height variable by $1$ if an arrow going into the vertex is crossed or decreases the height variable by one if an arrow going out of the vertex is crossed.  The configurations in \subref{8v} break the ice rule of the six-vertex model, and are  vortices of the height field.\label{fig:vertices} }
	\label{fig:baxter}
\end{figure}

\begin{figure}[hbt]
 	\includegraphics[width=.4\textwidth]{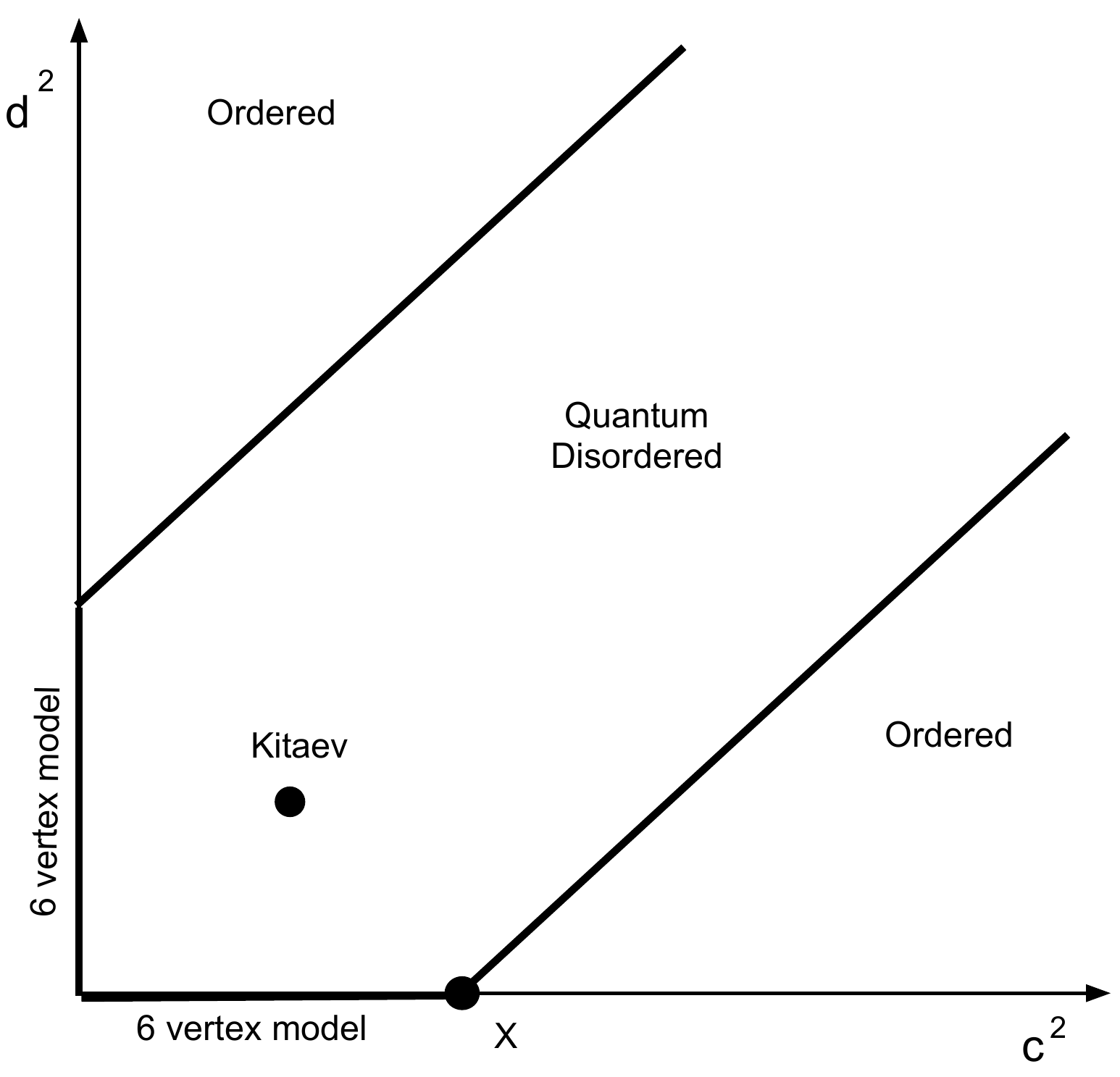}
	\caption{Phase diagram of the quantum eight vertex model (from Ref. [\onlinecite{Ardonne2004}]).
	Bold lines represent quantum critical lines.  The effective field theory along the lines labelled ``six-vertex model''  is the quantum Lifshitz model. Here $c$ and $d$ represent amplitudes of the wave function (see text). The quantum disordered phase is a $\mathbb{Z}_2$ topological phase and the point labelled ``Kitaev'' is the toric code. At the level of the wave function, the two critical lines are related by Kramers-Wannier duality. We will be studying the theory near the multicritical point, $X$, into the full eight vertex model along the critical line with $d^{2},c^{2} \neq 0$.}
	 \label{fig:phase}
 \end{figure}

An interesting system of this type is the quantum eight-vertex model of Ref.[\onlinecite{Ardonne2004}].  This model is a generalization of the 2D quantum dimer model. In this case the degrees of freedom are arrows placed on bonds of the square lattice. The allowed configurations are shown in Fig. \ref{fig:baxter}. This model also has an RK-type point at which the ground state wave function has local weights with the same form as the Gibbs weights in the corresponding Baxter (or eight-vertex) model on the square lattice, {\it i.e.} the Baxter wave function. The partition function for the classical model (which in the present context is the norm of the Baxter wave function) was solved by Baxter.\cite{Baxter-1982} The classical Baxter model has ordered and disordered phases separated by two critical lines (with continuously varying critical exponents) that meet at a multicritical point (see Figure \ref{fig:phase}).\cite{Kadanoff-1979} 

Ardonne {\it et al.}\cite{Ardonne2004} used the known results of the 2D classical Baxter model to determine the phase diagram of the quantum counterpart at its RK ``point'', shown in Fig. \ref{fig:phase}, as well as the behavior of the equal-time correlators in the phases and at (quantum) criticality. A key result of that work is that the disordered phase of the classical model is the $\mathbb{Z}_2$  topological phase of the quantum model. In particular there is  a special point deep inside this phase (where all the weights are equal)  at which the quantum Hamiltonian is identical to the Hamiltonian of Kitaev's toric code.\cite{Kitaev-2003} Hence, the Baxter wave function can be used to study a quantum phase transition from an ordered phase to a $\mathbb{Z}_2$   topological phase by crossing one of the critical lines.

However, the knowledge of the ground state wave function (and hence of all the equal-time correlation functions) is not sufficient to determine the universality class of the {\em quantum} phase transition. We therefore have the strange situation where much is known about the phases on either side of a critical point, but the theory at the critical point is incomplete. Although the quantum dimer models (and the quantum eight-vertex model) are not exactly solvable, variational arguments by Rokhsar and Kivelson\cite{RK1988} suggested that the dynamic critical exponent of these problems is $z=2$. This observation was formalized by Henley\cite{Henley1997} and in more detail by Moessner {\it et al.}\cite{Moessner-2001c} who suggested 
an effective field theory, obtained by coarse-graining the (dual) height representation, which has $z=2$. Ardonne {\it et al.}\cite{Ardonne2004} studied the consequences of this theory, which they dubbed the ``quantum Lifshitz model.'' They conjectured that this model represents  the universality class of generalized quantum dimer models and  both  critical lines of the quantum eight-vertex model.

The main purpose of this paper is to extend the analysis of Refs.[\onlinecite{Fradkin2004}] and [\onlinecite{Vishwanath-2004}] to include the effects of such magnetic (vortex) perturbations on the quantum Lifshitz model and to assess their effects on the  quantum universality class and more specifically on the dynamical scaling exponent $z$. We will specifically look at the case of the quantum eight-vertex model. Although the Baxter wave function obeys KW self-duality, the quantum Hamiltonian of the quantum-eight vertex model (whose exact ground state is the Baxter wave function) does not. This implies knowing the static properties of the correlators along the six-vertex model line it is possible to determine  what they are along the dual line, i.e. the line $d^{2},c^{2}\neq 0$ in Figure \ref{fig:phase}, using KW duality. However, since the full quantum Hamiltonian is not invariant under KW duality, we will not be able to infer the properties of the dynamical correlators along the dual line from their behavior along the six-vertex model line. In particular it is possible that the dynamical exponent $z$ may not be equal to $2$ along the dual line. More generally, we will show in this paper  that  along the dual line, the effective field theory (the quantum Lifshitz model)  is in general unstable under the action of perturbations that break the RK condition. Thus the dynamics of this system with a RK condition is very different than without it. 

Our strategy is to consider the quantum stability of the quantum Lifshitz model under the effects of generic local perturbations and to use the renormalization group to infer the behavior of the system.  To this end we include in the effective Lagrangian in 2+1 dimensions operators which describe the local creation and destruction (in space and time) of electric and magnetic charges, i.e. local fluctuations of dimers (or arrows in the Baxter language) and processes involving ``holons'' (or equivalently that break the symmetry of the Baxter model down to $\mathbb{Z}_2$). These local perturbations violate the RK condition. The recent Monte Carlo results of Isakov and coworkers,\cite{Isakov2011} which motivated our work, simulate the dynamics of a system in which the RK condition is obeyed. Marginally irrelevant perturbations and relevant operators which move the system away from quantum criticality are also included in the effective Lagrangian. The RG is constructed using an operator product expansion (OPE) adapted to anisotropic systems (with dynamical exponent $z>1$.) Consistently with the earlier results\cite{Grinstein1981,Fradkin2004} (and with the Monte Carlo simulations\cite{Isakov2011}) we find that the $z=2$ dynamics of the quantum dimer model (and the quantum six-vertex model) is (marginally) stable in the absence of ``magnetic'' (vortex) perturbations, but it becomes unstable if these perturbations are included. In this case the actual behavior of the system cannot be determined perturbatively. We conclude that the likely result is that either the (quantum) transition becomes first order or, if it is continuous, it is controlled by a non-perturbative fixed point (presumably with ``generic'' $z=1$ dynamics.) 

This paper is organized as follows. In Section \ref{sec:statement} we summarize the current status of the problem of the quantum dynamics of generalized quantum dimer models in 2D. In Section \ref{sec:qlm} we describe the quantum Lifshitz model in detail. In Section \ref{sec:RG} we present the one loop RG calculation performed, details of which (including the OPE) can be found in Appendix \ref{append:OPE} by the interested reader. In Section \ref{sec:disc} we discuss the differences which result from dynamics that preserve the RK condition (and in the Monte Carlo simulations of Ref.[\onlinecite{Isakov2011}] and more general dynamics resulting by perturbing the quantum Lifshitz model as we use here. Finally, in Section \ref{sec:conclusions} we discuss our
results and some open problems.

\section{Quantum Dynamics of Generalized Quantum Dimer Models}
\label{sec:statement}

The validity of the quantum Lifshitz model of the effective field theory of generalized quantum dimer models at their RK points was established in Refs. [\onlinecite{Moessner-2001c,Ardonne2004}]. Whether this theory describes the quantum dynamics away from RK points is far less obvious. In this section we summarize what is known about this issue. 

The validity of this conjecture has been considered by several authors. The combined effects of perturbations due to so-called ``charge'' operators (defined in later sections) and various non-linear gradients of the coarse-grained height field this quantum phase transition was studied using perturbative renormalization group (RG) methods.\cite{Fradkin2004,Vishwanath-2004} It was found that in many cases the RG flows destabilize the quantum Lifshitz fixed line, whether by rendering the quantum phase transition first order or by replacing it with an infinite sequence of ``tilting'' transitions (dubbed an incomplete devils' staircase).  Still in some cases the the quantum Lifshitz fixed line was found to be perturbatively stable (up to marginally irrelevant operators). This is what happens in the  quantum dimer model on the square lattice\cite{RK1988} and in the quantum eight-vertex model along the six-vertex model line 
(the $d=0$ axis of Fig. \ref{fig:phase}.)\cite{Ardonne2004,Castelnovo-2006,Papanikolaou2007b, Shannon} 
Nevertheless, these perturbative RG studies cannot determine (or exclude) the possibility that one unstable trajectory of the RG flow may drive the system into another fixed point (not perturbatively accessible) which would be, presumably, in the conventional $z=1$ universality class.

It is known that excitations that break the dimer constraints (by turn a bipartite lattice into a non-bipartite lattice)\cite{Moessner2001} or that break the ice-rule constraint of the six-vertex model (by  addition of  vertices with weight $d$ in the eight-vertex model\cite{Ardonne2004} (see Fig.\ref{fig:baxter})) trigger a transition into a $\mathbb{Z}_2$ topological phase. 
In terms of the dual height models (and hence of the quantum Lifshitz model) these perturbations are vortex (and hence ``magnetic'') operators. The effects of these magnetic perturbations on the quantum criticality of these systems is presently  poorly understood. What we know is that in the case of the quantum dimer models on the square lattice perturbations that break the bipartite property  of the lattice are relevant and drive the system immediately into a topological phase.\cite{Moessner2001,Ardonne2004} The same behavior is found along the six-vertex model line of the Baxter wave function.\cite{Ardonne2004} Using the Kramers-Wannier (KW) self-duality that the Baxter {\em wave function} has, Ardonne {\it et al.} were also able to compute the scaling behavior of the excitations along the ``dual'' critical line, the phase boundary between the topological phase and the ordered phase, along the critical line where $d^{2}\neq 0$  and $c^{2} \neq 0$ (see Figure \ref{fig:phase}). However these results do not determine what effects, if any, these excitations have on the dynamics and its scaling behavior.

While perturbations of the quantum Lifshitz theory has been a topic of interest to many, these studies so far have been limited to ``charged" excitations. Physically these perturbations can be thought of as point defects in the height model and are perturbations that preserve the ice rule (two arrows in and two arrows out at each lattice site) of the six vertex model (see Figure \ref{fig:vertices}). As such they are limited to the six-vertex limit of the full quantum eight vertex model. In the full quantum theory, magnetic vortex like excitations are also allowed. These correspond to topological defects to the height representation of the model. On the lattice, they are violations of the ice rule bayed by the allowed configurations (shown Figure \ref{fig:vertices}). Whereas ``charge" excitations can be handled by standard techniques, the effect of magnetic vortex excitations is less trivial and because of this, much is not known in the full quantum eight vertex model. It is the effect of these magnetic vortex excitations with the presence of charged excitations that forms the subject of this paper. We will examine the effect of these perturbations along the critical line within the full eight vertex model (see Figure \ref{fig:phase}).


Recent numerical studies by Isakov et. al. \cite{Isakov2011} have given further impetus for the study of the effects of magnetic excitations in the quantum eight-vertex model. In their study, they used a stochastic classical approach to simulate  the classical critical dynamics of the classical eight-vertex model. In this approach one does not directly simulate the dynamics of a quantum system in $2+1$ dimensions but instead, relies on strictly enforcing the RK condition, and does a classical simulation of the relaxation to equilibrium of a 2D system. In this approach the effects of magnetic excitations is taken into account while requiring the system to strictly obey the RK constraint.
Indeed, the classical stochastic theory is essentially a Langevin equation (or equivalently a master equation) for the classical stochastic time evolution, and has a path integral representation developed by Martin, Siggia and Rose.\cite{MSR} This theory is known to have a form of supersymmetry\cite{DeDominicis1978,ZJ1986,ZinnJustin} which ensures that the Langevin dynamics is strictly obeyed. In contrast the quantum Lifshitz model is defined by a quantum mechanical action in $2+1$ dimensions.

At the level of the free quantum Lifshitz models (\emph{i.e.} along the six vertex line) the stochastic classical simulation and the quantum mechanical path integral are essentially equivalent to each other.\cite{Moessner-2001c} However, they do not necessarily agree once the perturbations are included. Indeed in the Langevin approach one perturbs the wave function, whereas in the $2+1$ dimensional quantum field theory one perturbs the full quantum Hamiltonian whose form is no longer the sum of projection operators and, hence, in general is not equivalent to a classical Langevin dynamics problem. Isakov and coworkers carried out a detailed classical Langevin simulation of the eight-vertex model wave function. Along the six-vertex model line, $d=0$, their results are consistent with $z=2$ along the entire fixed line, in agreement with earlier analytic results\cite{Grinstein1981,Fradkin2004,Vishwanath-2004} and with the results that we report here.
However, for $d>0$ Isakov and coworkers find the startling result that the dynamic exponent $z$ varies (and apparently in a continuous fashion) 
along the phase boundary between the disordered, topological phase and the ordered phase. 
As a check they confirmed that at the special point where the classical eight-vertex model becomes two decoupled classical Ising models, 
the classical dynamic exponent is $z\sim 2.196$ (which is consistent with the known value of $z$ for the classical 2D Ising model). 
Thus, these classical Monte Carlo simulations imply that the quantum theory is unstable and that the dynamics of the effective field theory is not $z=2$ as it is along the six-vertex line.

These results motivated us to study the full quantum Lifshitz action and the effect of perturbations as one moves away from the six-vertex line. We do this by perturbing directly the quantum Lifshitz model action with both local charge and magnetic excitations. 
We should note that this approach does not enforce the RK condition and hence it is not equivalent to the classical simulations of Isakov and coworkers. 
By an analysis of the RG equation we will see that  the perturbed quantum Lifshitz model has one line of fixed points. We find that the dynamic critical exponent is $z=2$ along this line and it does not change.  On the other hand, the magnetic excitations are described by a marginally relevant perturbation. Thus this theory has two marginally relevant operators whose coupling constants grow monotonically. Here two things can happen: one option is that the quantum phase transition becomes first order, in which case there is no quantum criticality. The other option is that the system flows to a new fixed point, which is not perturbatively accessible. Presumably this new fixed point will be have  ``generic'' $\mathbb{Z}_2$ quantum phase transition with dynamic exponent $z=1$. 
Clearly, our one-loop RG cannot be used to construct this theory. 

In the regime in which our RG calculation is reliable, it is consistent with a dynamical exponent  $z=2$. On the other hand, in order to reproduce the simulation done by Isakov and coworkers along the dual critical line of the classical Baxter model would require to  construct a Martin-Siggia-Rose supersymmetric effective action which  must have an exactly marginal operator to describe the equilibrium  spatial correlations (with varying exponents) and with a continuously changing $z$. Although to the best of our knowledge this theory does not yet exist, we can speculate on a possible scenario to understand the results of Isakov {\it et al.} It is well known from the work of Kadanoff\cite{Kadanoff-1979b} that the critical line of the Baxter model dual to the six vertex model line can be mapped to the partition function of a Gaussian model perturbed by electric and magnetic charges. The problem that we studied in this paper is the quantum analog of this problem. Now, a Langevin equation constructed to study the classical critical dynamics of this problem will be local in the scalar field $\phi$ of the 2D classical gaussian model. Since the electric charges are represented by local vertex operators of the Gaussian model, these excitations are also local in the stochastic process described by the Langevin equation. However the magnetic charges (or vortices), which are dual to the electric charges, are non-local in terms of the field $\phi$ which represents the height variable. They are physically edge defects in the lattice picture. This suggests that the relaxation processes become increasingly non-local as the coupling to the magnetic excitations increases. Thus it is conceivable that this non-locality may give rise to a varying $z$. 
However,  even if these considerations are correct, a continuously varying $z$ is inconsistent with a theory with a local action with an exact marginal operator.

In the next section we provide details of the quantum Lifshitz model for the reader unfamiliar with the subject matter. In section IV, we will then return to this question of flowing dynamical exponent and present details of our analysis. Detailed calculations can be found in Appendix B.


\section{Quantum Lifshitz Theory}
\label{sec:qlm}

The quantum Lifshitz model is the simplest continuum field theory that has been proposed to describe the quantum six vertex and quantum dimer models on the lattice. The simplest of illustration of these ideas can be seen with the quantum dimer model on the square lattice which 
acts on the configuration space 
of a classical model of close packed (one dimer per vertex) dimers on a two dimensional lattice. The same considerations apply to the quantum eight vertex model that we will discuss below, briefly. For a much more thorough discussion relating the quantum Lifshitz theory and the six vertex model, we refer the reader to Ref. [\onlinecite{Ardonne2004}].

\subsection{Quantum Dimer Models}

Let $\{ \vert \alpha \rangle\}$ represent a set of linearly independent and orthonormal states which in the dimer models are the dimer coverings of the lattice and in the eight vertex models are the allowed arrow configurations. The quantum Hamiltonians are local hermitian operators that act on this configuration space.
The the kinetic energy term describes a quantum mechanical process mapping two configurations that differ by local moves (flips) while the potential energy  term of the Hamiltonian penalizes each flippable plaquette $i$ of the lattice. It is conventionally written in terms of operators acting on all the plaquettes of the lattice 
\begin{equation}
	\mathcal{H} = \sum_{i} \left(\hat{V}_{i} - \hat{F}_{i}\right)
\end{equation}
Here $F_i$ is the local kinetic energy term (the flip operators) and $V_i$ are the local potential energy terms.
For special choices of parameters each term of this Hermitian operator obeys the RK condition\cite{RK1988},  $(\hat{V}_{i}-\hat{F}_{i})^{2}=2 (\hat{V}_{i}-\hat{F}_{i})$.
Hence, the Hamiltonian can be written as $\mathcal{H}=\frac{1}{2}\sum_{i} Q_{i}^{\dagger}Q_{i}$ where $Q_{i}=(\hat{V}_{i}-\hat{F}_{i})$. Hamiltonians of this form necessarily have eigenvalues $E\geq 0$ and more importantly, if one can find a state annihilated by $\hat{Q}_{i}$, then one has found the ground state. The equal amplitude sum over all possible classical dimer coverings is one such state. Then, the properly normalized ground state wave function is given by,
\begin{equation}
	\vert \psi_{0} \rangle = \frac{1}{\sqrt{Z} } \sum_{\alpha} \vert \alpha \rangle.
	\label{eq:RVB}
\end{equation}
$Z$ is simply number of all dimer configurations on some finite two dimensional lattice and is precisely the classical partition function of two dimensional dimers. 

Ardonne {\it et. al} generalized this construction to cases in which the wave function is not an equal-amplitude superposition as in the ``RVB'' wave function, Eq.\eqref{eq:RVB}, but to one in which the configurations are assigned a {\em local} weight. In the case of the quantum eight-vertex model along the RK lines  the wave function is\cite{Ardonne2004}
\begin{equation}
\vert \Psi_{q8v}\rangle=\sum_{\{ \mathcal{C}\} } a^{N_a[\mathcal{C}]}b^{N_b[\mathcal{C}]}b^{N_a[\mathcal{C}]}c^{N_c[\mathcal{C}]}d^{N_d[\mathcal{C}]} \vert\mathcal{C}\rangle
\label{eq:Psiq8v}
\end{equation}
where $\{ \mathcal{C} \}$ are the configurations of the eight-vertex model with the weights shown in Fig. \ref{fig:baxter}.
The norm (squared) of this wave function, $||\psi_{q8v}||^2$, is equal to the partition function of the classical 2D eight-vertex (Baxter) model with weights $a^2$, $b^2$, $c^2$, and $d^2$. In Figs. \ref{fig:baxter} and \ref{fig:phase} we show the vertices and the resulting phase diagram represented by this wave function (for the case $a=b=1$) in terms of $c^2$ and $d^2$. For similar expressions for generalized dimer models see Refs.[\onlinecite{Castelnovo-2006,Alet-2006,Papanikolaou2007b}].

\subsection{Mapping Quantum Dimer and Vertex Models to the Quantum Lifshitz Model}

This analysis extends to the computation of equal time correlation functions in the ground state, and one finds that these correlators are simply given by correlation function of the classical two dimensional theory. \cite{RK1988,Ardonne2004} Hence, for $\mathcal{H}_{RK}$ on the square lattice, the equal time static correlation functions are algebraically decaying and $\mathcal{H}_{RK}$ is said to sit between two ordered phases of the dimers. On the triangular lattice, however, the equal time correlation functions in the ground state are instead exponentially decaying and the quantum dimer model is gapped. The system is in fact in a topological phase.\cite{Moessner2001}

The dimer configurations (as well as the six vertex configurations) can be mapped to a height variable on the dual lattice.\cite{Nienhuis1987} This mapping holds for both the classical and the quantum problem. The dimer-height mapping is exact provided the dimer configurations are fully packed, i.e. every site of the lattice belongs to one and only one dimer, and that the lattice is bipartite. The violation of either condition is mapped in the height model into a topological defect which is referred to as a vortex.\cite{Nienhuis1987} The same issue arises in vertex models in which the vertices which violate the local conservation law of the six vertex model are also represented  by vortices in the (dual) height model picture.

Let us assume for now that no such topological singularities are present. We can now use a coarse-grained picture of the (dual) height model.\cite{Nienhuis1987, Kondev-1996} In this effective continuum limit, the coarse-grained height variable becomes a free boson $\varphi(x,t)$ (see Figure \ref{fig:vertices}) with compactification radius given by the modulo $4$ condition on the lattice. A continuum Hamiltonian conjectured by Henley\cite{Henley2004,Henley1997} and expanded by Moessner and collaborators\cite{Moessner-2001c} to belong in the same universality class as the square lattice quantum dimer model is,
\begin{equation} \label{eq:qlm}
	\mathcal{H} = \int d^{2}x \, \, \left[ \frac{\Pi^{2} }{2} + \frac{\ka^{2} }{2} (\nabla^{2}\varphi)^{2} \right],
\end{equation}
where $\Pi = \dot{\varphi}$ as usual. The associated Euclidean action is given by,
\begin{equation} \label{eq:qLt}
	S = \int d^{3}x \,\, \left[ \frac{\dot{\varphi}^{2} }{2} + \frac{\ka^{2} }{2} (\nabla^{2}\varphi)^{2} \right] .
\end{equation}
This action also arose in the study of three dimensional classical statistical mechanics in the study of Lifshitz points in smectic liquid  crystals \cite{Grinstein1981} and hence the theory was dubbed the quantum Lifshitz model.\cite{Ardonne2004} More importantly, the ground state wave function is related to a two dimensional Euclidean free boson and represents a line of critical points parameterized by the constant $\kappa$.\cite{Grinstein1981}

This can be done by quantizing the Hamiltonian by imposing the equal time commutation relations,\cite{Ardonne2004}
\begin{equation}
	\left[ \varphi(\vec{x}),\Pi(\vec{y}) \right] = i \delta^{(2)}(\vec{x}-\vec{y}),
\end{equation}
so that in the Schr\"odinger picture the canonical momentum can be written as $\Pi(\vec{x}) = i \delta / \delta\varphi(\vec{x})$. Schr\"odinger equation can be written as,
\begin{equation}
	\int d^{2}\vec{x} \,\, \left[ - \frac{1}{2} \left( \frac{\delta}{\delta\varphi } \right)^{2} + \frac{\ka^{2}}{2} \left(\nabla^{2}\varphi \right)^{2} \right] \Psi_{0}[\varphi] = E\Psi_{0}[\varphi].
\end{equation}
Now, the Hamiltonian can be written as a self-adjoint operator as in the RK Hamiltonian. Defining the operators
\begin{eqnarray}\label{eq:qlmops}
	Q(\vec{x}) &=& \frac{1}{\sqrt{2} } \left( \frac{\delta}{\delta\varphi} + \ka \nabla^{2}\varphi \right) \nonumber \\
	&& \\
	Q^{\dagger}(\vec{x}) &=& \frac{1}{\sqrt{2} } \left( -\frac{\delta}{\delta\varphi} + \ka \nabla^{2}\varphi \right) \nonumber.
\end{eqnarray}
The normal ordered quantum Hamiltonian is then given by,
\begin{eqnarray} \label{eq:qdmHam}
	\mathcal{H} &=& \frac{1}{2} \int d^{2}\vec{x} \,\, \bigg\{ Q^{\dagger}(\vec{x}),Q(\vec{x}) \bigg\} - \epsilon_{vac} V \nonumber \\
	&=& \int d^{2}\vec{x} \,\, Q^{\dagger}(\vec{x})Q(\vec{x}),
\end{eqnarray}
where $V$ is the area and $\epsilon_{vac}$ is the UV divergent vacuum energy,
\begin{equation}
	\epsilon_{vac} = -\frac{\ka}{2} \lim_{\vec{y}\rightarrow \vec{x} } \nabla^{2}_{\vec{x} } \delta^{(2)}(\vec{x}-\vec{y}).
\end{equation}
The normal ordered Hamiltonian is a self-adjoint operator with energy $E\geq0$ and therefore any state for which $Q\Psi_{0}[\varphi]=0$ must be the ground state. This is simply a first order functional differential equation which is easily solved by,
\begin{equation}\label{eq:qdmPsi}
	\Psi_{0}[\varphi] = \frac{1}{\sqrt{Z} } e^{-\frac{\ka}{2} \int d^{2}\vec{x} \, \, \left( \nabla\varphi\right)^{2} },
\end{equation}
where $Z$ is the normalization factor,
\begin{equation}
	Z = \int [D\varphi] e^{-\ka\int d^{2}\vec{x} (\nabla\varphi)^{2}}.
\end{equation}

Correlation functions at equal time in the ground state are 
\begin{eqnarray}
	&&\langle \textrm{vac} \vert \mathcal{O}[\varphi(\vec{x}_{1} )] \dots \mathcal{O}[\varphi(\vec{x}_{n} )] \vert \textrm{vac}\rangle =\nonumber \\
 	&& \int [D\varphi] \,\, \mathcal{O}[\varphi(\vec{x}_{1} )] \dots \mathcal{O}[\varphi(\vec{x}_{n} ]\,\,  e^{-\ka \int d^{2}\vec{x} \,\, (\nabla\varphi)^{2} },
\end{eqnarray}
and are hence directly related to the power law correlation functions of the free boson field theory. 
The behavior of the correlation functions of this theory is controlled by the parameter $\kappa$ of the quantum Lifshitz Hamiltonian. As in the classical 2D XY-model,\cite{Kadanoff-1979b} and in the Luttinger model of 1D fermionic systems, there is a line of critical points characterized by the parameter $\kappa$. The parameter $\kappa$ is related in a non-universal way to the microscopic models. For instance, Ardonne {\it et. al} gave an explicit expression for the relation between $\kappa$ and the weights $c$ and $d$ in the quantum eight-vertex model.\cite{Ardonne2004}

In particular, the operator 
\begin{equation}
	\mathcal{O}_{q}[\varphi(\vec{x})] = e^{i q \varphi(\vec{x})}
\end{equation} 
for $q\in \mathbf{Z}$ creates a bosonic coherent state which we call a charge excitation. This operator enters in the expression for the order parameters of the ordered phases of the quantum dimer model\cite{Fradkin2004} and of the quantum eight-vertex model.\cite{Ardonne2004}  This correlation function obeys a power-law of the form
\begin{equation}
	\langle \mathcal{O}_{q}(\vec{x}) \mathcal{O}_{-q}(\vec{y}) \rangle \simeq \frac{1}{\vert \vec{x}-\vec{y} \vert^{q^{2}/4 \pi \ka} }
\end{equation}
which follows from a similar set of manipulations performed in the XY-model and using the quantities in Appendix \ref{append:OPE}. It follows that the scaling dimension for the operator $\mathcal{O}_{q}$ is 
\begin{equation}
	\Delta_{q} = \frac{q^{2}}{8 \pi \ka}.
\end{equation}

The ``magnetic charge'' (vortex) operator  is given by\cite{Ardonne2004}
\begin{equation} \label{eq:dual}
 	\tilde{\mathcal{O}}_{m}(\vec{x}) = e^{i 2 m \int d^{2}\vec{z} \,\,  \arg(\vec{x}-\vec{z}) \, \Pi(\vec{z}) }.
\end{equation}
The action of this operator on the ground state is to shift in the boson configuration to another with a vortex singularity at $\vec x$ with topological charge (vorticity) $m$.  In other words, it introduces a jump discontinuity in the height variables and corresponds to an addition of  ``source" and ``sink" vertices to the six vertex model to give the full eight vertex model (see Figure \ref{fig:vertices}). \cite{Ardonne2004} The defect (Dirac string) associated with these magnetic vortices corresponds to the non-local term $\exp \left( \int dz \arg(x-z) \Pi(z) \right)$. It amounts to a singular gauge transformation on the field $\varphi$. Its action on the ground state wave function can be written as,
\begin{equation}
	\langle \tilde{\mathcal{O}}_{m_{1}}(\vec{x}_{1} )\dots \tilde{\mathcal{O}}_{m_{n}}(\vec{x}_{n}) \rangle = \frac{1}{Z} \int D\varphi \, \, e^{-\ka \int d^{2}\vec{x} \, \, (\nabla\varphi + \vec{A} )^{2} }
\end{equation}
where $\vec{A}$ satisfies,
\begin{equation}
 	\epsilon^{ij} \nabla_{i}A_{j} = 2\pi \sum_{\ell=1}^{n} m_{\ell} \,\, \delta^{2}(\vec{z} - \vec{x}_{\ell})  \label{eq:As}
\end{equation}
This expression is non-vanishing only if the charge-neutrality condition,  $\sum_{\ell} m_{\ell}=0$, is obeyed. This condition  reflects the fact that the wave function of the quantum Lifshitz model is vortex-free. The same considerations apply to the electric charges.

Eq.\eqref{eq:As} has the solution
\begin{equation}
A_{j} = \sum_{\ell} m_{\ell}\, \partial_{j}\, \textrm{arg}(\vec{z}-\vec{x}_{\ell})
\end{equation}
By using the Cauchy-Riemann condition, 
\begin{equation}
\partial_{j}\,\, \textrm{arg}(\vec{z}-\vec{x}_{\ell}) = \epsilon_{ij} \partial^{i} \log \vert \vec{z}-\vec{x}_{\ell} \vert
\end{equation}
 its straight forward to find the correlation function. Specializing to the two point function, the result is,
\begin{equation} \label{eq:vort_corr}
	\langle \tilde{\mathcal{O}}_{m}(\vec{x} ) \tilde{\mathcal{O}}_{-m}(\vec{y})  \rangle\simeq  \left( \frac{1}{\vert\vec{x}-\vec{y}\vert} \right)^{4\pi\kappa m^{2} }
\end{equation}
as expected, the equal time critical exponent is given by, $\tilde{\Delta}_{m} = 2\pi \ka m^{2}$.

From this close connection with conformal field theories, its easy to see that the static behavior of the conformal quantum critical points and specifically the quantum Lifshitz theory are well understood. The \emph{dynamic} behavior, however, is another story and has remained a largely unexplored. In recent numerical computations based on simulations of the Langevin equation, it was proposed that by perturbing the Lifshitz theory by allowing for line defects (i.e. the operator $\tilde{\mathcal{O}}_{m}$) it would be found that the dynamical exponent would flow continuously from $z=2$ to $z=1$. If true, this would contradict much of what is known. In most models the dynamical exponent is found to jump discontinuously between integer values.\cite{CL,Hohenberg-Halperin} However, because of the close relationship between the quantum Lifshitz theory and the XY-model where all critical exponents vary continuously, it was hypothesized that perturbations in the quantum Lifshitz theory would result in a flowing dynamical exponent. \cite{Isakov2011} In the remainder of the paper, this issue is explored in further detail.

\section{Perturbative Renormalization}
\label{sec:RG}

In this section we will consider a system that consists of a $2+1$-dimensional quantum Lifshitz model plus a set of  local electric and magnetic perturbations. Such a generic perturbed system does not (in general) respect an RK condition and therefore it is not equivalent to a classical system with a partition function perturbed by similar operators. In other terms these two problems are not in general equivalent as the $2+1$-dimensional theory in general is not consistent with the supersymmetry implied by the Langevin equation.

It has been proposed that the square lattice quantum dimer model is described by the pure quantum Lifshitz theory.\cite{Henley1997,Moessner-2001c,Ardonne2004} We wish to study the theory near the six vertex model and perturb into the full eight vertex model by allowing for magnetic vortices which break the ice rule of the six vertex model. This point is Kosterlitz-Thouless (KT) like and to move deeper into the ordered phase corresponds to varying the fugacity of the vertex operators. The theory along the quantum six vertex line is analogous to the sine-Gordon model and was studied in detail by Grinstein \cite{Grinstein1981} where it was called an anisotropic sine-Gordon model. Specifically we consider one set of perturbations,
\begin{equation}
	\frac{\vertex}{2a^{4} } \left( \mathcal{O}_{q}(\vec{x},t) + \mathcal{O}_{-q}) (\vec{x},t)\right) = \frac{\vertex}{a^{4} } \cos\left( q\varphi(\vec{x},t) \right)
\end{equation}
As noted above, the equal time scaling dimension of this operator is $\Delta_{q} = q^{2}/(8\pi\ka)$.

Secondly, we need to perturb the quantum six vertex model by allowing the ice rule to be broken. This corresponds to an operator, 
\begin{equation}
	\frac{\dvertex}{2a^{4} } \left( \tilde{\mathcal{O}}_{m}(\vec{x},t) + \tilde{\mathcal{O}}_{-m}(\vec{x},t) \right) 
\end{equation}
where the time dependent vortex operator is given by,
\begin{eqnarray}
	&& \tilde{\mathcal{O}}_{m}(\vec{x}_{\ell},t_{\ell}) = \\
	&& \textrm{exp} \left( i 2 m \int_{-\infty}^{\infty} d\tau \int d^{2}\vec{z} \, \, \arg(\vec{z}-\vec{x}_{\ell} ) \delta(\tau-t_{\ell}) \Pi(\vec{z},\tau) \right).\nonumber 
\end{eqnarray}
This operator gives the same equal time correlation functions (\ref{eq:vort_corr}) and its equal time scaling dimension is given by $\tilde{\Delta}_{m} = 2\pi  \ka m^{2} $. Analogous to the anisotropic sine-Gordon model,\cite{Grinstein1981} the coupling constant at the KT-point is $\kappa_{c}=1/(32\pi)$, and one easily finds that the $\mathcal{O}_{q}$ is marginal for $q=1$ and $\tilde{\mathcal{O}}_{m}$ is marginal for $m=8$.

Taking the effective low energy theory to be the quantum Lifshitz theory, we consider the action, 
\begin{eqnarray}
	\mathcal{S} &=&  \int d^{2}\vec{x} dt \,\, \frac{1}{2} (\partial_{t}\varphi)^{2} + \frac{\ka^{2}_{c} }{2}(\nabla^{2}\varphi)^{2} \nonumber \\
	&+& A (\nabla \varphi)^{2} +\delta (\nabla^{2}\varphi)^{2}+ u (\nabla\varphi)^{4}\nonumber \\
	& +& \frac{\vertex}{2a^{4} } (\mathcal{O}_{q} +\mathcal{O}_{-q}) + \frac{\dvertex}{2 a^{4} } (\tilde{\mathcal{O}}_{m} + \tilde{\mathcal{O}}_{-m} )
\end{eqnarray}
We expand about the fixed point where both $\tilde{\mathcal{O}}_{m}$ and $\mathcal{O}_{q}$ are marginal at $\ka_{c} = 1/32\pi $. Defining $\delta$ to be the distance away from marginality so that $\delta\ll1$, we look at the quadruple expansion in the small parameters $\delta, u, \alpha$ and $\tilde{\alpha}$ while treating $A=A(u,\alpha, \tilde{\alpha},\delta)$ as a renormalization group parameter; one fixes $A=0$ to give the Lifshitz theory. In this sense it is a parameter that constrains the flow equations properly as opposed to a coupling constant that flows under the RG. While $u$ is marginally irrelevant, the justification for including such an operator is that it is generated under renormalization anyway.

\subsection{Renormalization group for the perturbed quantum Lifshitz model}

Now, we want to study the flow of the various coupling constants when the variables $u,\alpha,\tilde{\alpha}$ and $\delta$ are small. This is done through a perturbative renormalization scheme. In such a procedure, one begins with an effective low energy field theory described by an action $S_{0}$ and a set of operators $\mathcal{O}_{i}$ with coupling constant $g_{i}$ and scaling dimension $\Delta_{i}$ that will be treated perturbatively. Treating the coupling constants as small, one can expand in powers of $g_{i}$,
\begin{eqnarray}
	\mathcal{Z} &=& \textrm{Tr } e^{-\beta \mathcal{H}_{0} - \int d^{3}\vec{x} \sum_{i} g_{i}\mathcal{O}_{i}(\vec{x}) } \nonumber \\
	&\sim& Z_{0} \Big[ 1 - \int d^{3}\vec{x} \, \sum_{i} g_{i}\langle \mathcal{O}_{i} \rangle \nonumber \\
	&+& \frac{1}{2! } \int d^{3}\vec{x} d^{3}\vec{y} \sum_{i,i'} g_{i}g_{i'} \langle \mathcal{O}_{i}(\vec{x}) \mathcal{O}_{i'}(\vec{y}) \rangle +\dots \Big] \label{eq:pert}
\end{eqnarray}
where the correlation functions are taken with respect to the low energy effective field theory. The renormalization is then performed by rescaling the short distance cutoff $a\rightarrow \lambda a$ where $\lambda= (1+\delta)$ where $\delta>0$ and asking how the couplings rescale so that the partition function is preserved. In the first term, this rescaling is simple. The coupling constant $g_{i}$ rescales as,
\begin{eqnarray}
	g_{i} \rightarrow \lambda^{d-\Delta_{i} } g_{i} \sim g_{i} + (d-\Delta_{i}) g_{i} \delta
\end{eqnarray}
In the second, one may use the operator product expansion. Under a rescaling, the limits of integration can be written as 
\begin{eqnarray}
	\int_{\vert \vec{x}-\vec{y} \vert >(1+\delta)a } =\int_{\vert \vec{x} -\vec{y} \vert > a } - \int_{a<\vert \vec{x}-\vec{y}\vert<a(1+\delta) }
\end{eqnarray}
The first term only gives the original contribution to the Hamiltonian. In the second, one may apply the operator product expansion. We know that when  two operators are nearby, we know that they can be thought of as effectively ``fusing'' into another operator in the theory.
\begin{eqnarray}
	& \lim_{\vert \vec{x} - \vec{y}\vert\rightarrow a}&  \langle \mathcal{O}_{i}(\vec{x}) \mathcal{O}_{i'}(\vec{y}) \rangle  \nonumber \\
	&=& \sum_{j} C^{j}_{i,i'}(\vec{x}-\vec{y}) \left\langle \mathcal{O}_{j}\left(\frac{\vec{x}+\vec{y}}{2} \right) \right\rangle
\end{eqnarray}
where $C^{j}_{i,i'}$ are the fusion coefficients.

To find the fusion coefficients, one looks at the three point function. Its a generic feature of $d$-dimensional critical systems that the three point correlation functions are restricted by translation, rotational and scale invariance to have the form
\begin{eqnarray}
	&& \langle O_{i}(\vec{x}_{1}) O_{j}(\vec{x}_{2}) O_{k}(\vec{x}_{3}) \rangle \simeq \nonumber \\
	&&  \frac{1}{ (\Delta_{12} )^{y_{i}+y_{j}-y_{k} }(\Delta_{13})^{y_{i}+y_{k}-y_{j} } (\Delta_{23})^{ y_{j}+y_{k}-y_{i} } },\nonumber \\
\end{eqnarray}
where $\Delta_{ij} = \vert \vec{x}_{i}-\vec{x}_{j} \vert$ and $y_{i}$ is the scaling dimension of $O_{i}$. The added complication in anisotropic fixed points is that scaling invariant functions of $s=\vert \vec{x}\vert^{2}/t$ may also appear, but the structure remains the same. Defining $s_{ij} = \Delta_{ij}^{2}/t_{ij}$, generically what we find that the three point function are scaling functions of the form,
\begin{eqnarray}
	&& \langle O_{i}(\vec{x}_{1}) O_{j}(\vec{x}_{2}) O_{k}(\vec{x}_{3}) \rangle \simeq \nonumber \\
	&& \frac{f(s_{12} ) f(s_{13}) f(s_{23}) }{ (\Delta_{12} )^{y_{i}+y_{j}-y_{k} }(\Delta_{13})^{y_{i}+y_{k}-y_{j} } (\Delta_{23})^{ y_{j}+y_{k}-y_{i} } }\nonumber \\
\end{eqnarray}
Now, in the limit that $\Delta_{12} \ll \Delta_{13}$ and $\Delta_{12} \ll\Delta_{23}$ and analogously for $t_{ij}$, the three point correlation function factors. If $2 \vec{R} = x_{1}+x_{2}$ and $2 \vec{r}=x_{1}-x_{2}$ (and likewise for time) while further setting the dummy variable $\vec{x}_{3}=0$, this can be written as, 
\begin{equation}
	\langle O_{i}(\vec{x}_{1}) O_{j}(\vec{x}_{2}) O_{k}(0) \rangle \simeq  c_{ijk}  \frac{f(s_{r}) }{r^{y_{i}+y_{j}-y_{k} } } \frac{f^{2} (s_{R}) }{R^{2 y_{k} } }.
\end{equation}
where the $c_{ijk}$ are numbers and the functions $f(s_{r})$ are scale invariant functions of $s_{r} = r^{2}/t$. Because of the more complicated scaling functions present, less divergent terms in $R$ are also present, but we keep only the leading divergent behavior here. To connect with the results in Grinstein\cite{Grinstein1981}, we finally want to integrate all values of $s$ to some fixed value $s_{0}$. Then, by taking the time to be small at this point, one integrates $0<s<\infty$. In addition, exchanging the time variable for a dimensionless $s$ variable, one needs to only consider terms which are divergent as $r^{-4}$ in the limit $r\rightarrow 0$.

By doing so, one finds that the contribution coming from the terms second order in the coupling constants, yields a contribution,
\begin{eqnarray}
	&-\frac{1}{2}&  \sum_{j}\frac{ c^{j}_{i,i'}  }{a^{\Delta_{i}+\Delta_{i'}-\Delta_{j} } } \int ds_{r} d\Omega f(s_{r})\nonumber \\
	&\times&  \int_{a<\vert \vec{x}-\vec{y}\vert<a(1+\delta) }d^{3}\vec{x}d^{3}\vec{y} \,\,\left\langle \mathcal{O}_{j}\left(\frac{\vec{x}+\vec{y}}{2} \right) \right\rangle .
\end{eqnarray}
As mentioned previously one integrates over all possible values of the aspect ratio and the solid angles. Provided that the integrals converge, this is a legitimate procedure and yields a simple number. Re-exponentiating the result, one finds that the coupling constant then flows as, 
\begin{equation}
	\frac{dg_{j} }{d \delta } = (d-\Delta_{j}) g_{j} - \frac{1}{2} S_{d} \sum_{i,i'} c^{j}_{i,i'} g_{i}g_{i'}.
\end{equation}

Hence, to find the $\beta$-functions various operator fusion coefficients $c^{j}_{i,i'}$ must be computed. This is outlined in Appendix B. Defining the quantity,
\begin{equation}
	J_{i} =\int_{0}^{\infty} ds \,\, s^{-2+2i} \textrm{exp}\left( 4 \int_{0}^{s/4} \frac{e^{-x}-1}{x} dx \right) 
\end{equation}
so that numerically, $J_{1} = 0.290$ and $J_{2}=0.111$, one finds that for the quantum Lifshitz model, the non-zero operator product expansion coefficients are, 
\begin{widetext}

\begin{eqnarray}
	:(\nabla\varphi)^{4}(\vec{x}): :(\nabla\varphi)^{4}(\vec{y}): &\simeq& \frac{1}{\pi^{3} } \log\left(\frac{32}{27} \right) \frac{\scale}{a^{4}} :\mathbf{1}: - \frac{4}{\pi^{2} }\left[ \frac{2}{12} - \log\left(\frac{4}{3} \right) \right]  \frac{\scale}{a^{2} } :(\nabla\varphi)^{2}(\vec{R}):  + \frac{41}{16 \pi } \scale :(\nabla\varphi)^{4}(\vec{R}): \nonumber \\
	&+& \frac{1}{\pi^{2} }\left[ - \frac{5}{12} +  \log \left(\frac{4}{3} \right) \right]  \scale \,\,  :(\nabla^{2}\varphi)^{2}(\vec{R}): \nonumber \\
	\mathcal{O}_{1}(\vec{x}) \mathcal{O}_{-1 }(\vec{y}) &\simeq& \frac{1 }{4} \frac{e^{4\gamma} }{4^{7} } \frac{J_{2} }{\pi}  \frac{\scale}{a^{4} } :\mathbf{1} : -  \frac{e^{4\gamma} }{4^{4} } \frac{J_{2}}{32}  \frac{\scale}{a^{2} } \,\,  :(\nabla\varphi)^{2}(\vec{R}): +   \frac{e^{4\gamma} }{4^{6} } \pi J_{2}  \scale \,\, :(\nabla\varphi)^{4}(\vec{R}): \nonumber \\
	& -&   \frac{e^{4\gamma} }{4^{6} } J_{1} \scale \, \, :(\nabla^{2}\varphi)^{2}(\vec{R}):  \nonumber  \\
	\tilde{\mathcal{O}}_{8}(\vec{x} )\tilde{\mathcal{O}}_{-8} (\vec{y})&\simeq& \frac{1}{4} \frac{e^{4\gamma} }{4^{7} } \frac{J_{2} }{\pi}  \frac{\scale}{a^{4} } :\mathbf{1} : -  \frac{e^{4\gamma} }{4^{4} } \frac{J_{2}}{32}  \frac{\scale}{a^{2} } \,\,  :(\nabla\varphi)^{2}(\vec{R}): + \frac{e^{4\gamma} }{4^{6} } \pi J_{2}  \scale \,\, :(\nabla\varphi)^{4}(\vec{R}): \nonumber \\
	& -&  \frac{e^{4\gamma} }{4^{6} } J_{1} \scale \, \, :(\nabla^{2}\varphi)^{2}(\vec{R}): \nonumber  \\
	\mathcal{O}_{1}(\vec{x}) :(\nabla\varphi)^{4}(\vec{y}): &\simeq& \frac{1 }{2^{7}\pi } \frac{16}{\pi }\log\left(\frac{32}{27}\right) \scale \, \, : \mathcal{O}_{-1}(\vec{R}): \nonumber \\
	 \mathcal{O}_{1}(\vec{x}) : (\nabla^{2}\varphi)^{2}(\vec{y}) : &\simeq& -\frac{1 }{32 \pi} \scale \,\, : \mathcal{O}_{-1}(\vec{R}): \nonumber \\
	 \tilde{\mathcal{O}}_{8}(\vec{x}) :(\nabla\varphi)^{4}(\vec{y}): &\simeq& \frac{1 }{2^{7}\pi } \frac{16}{\pi }\log\left(\frac{32}{27}\right) \scale \, \, : \tilde{\mathcal{O}}_{-8}(\vec{R}): \nonumber \\
	 \tilde{\mathcal{O}}_{8}(\vec{x}) : (\nabla^{2}\varphi)^{2}(\vec{y}) : &\simeq& -\frac{1 }{32 \pi} \scale \,\, : \tilde{\mathcal{O}}_{-8 }(\vec{R}): 
	 \label{eq:one-loop-RG}
\end{eqnarray}
\end{widetext}

From the operator product expansion and fusion coefficients, one can easily find the $\beta$-functions for the various coupling constants. The renormalization group equations are similar to those of the two dimensional isotropic sine-Gordon model. \cite{Grinstein1981, Nienhuis1984,Boyanovsky1989}. The $\beta$-functions have the form, 
\begin{eqnarray} \label{eq:beta_functions1}
	\beta(u) &=& - \frac{41}{32\pi } u^{2}  - \frac{\pi e^{4\gamma}}{4^{7} } J_{2} \left( \alpha^{2} -  \tilde{\alpha}^{2} \right) ,\nonumber \\
	\beta(\alpha)&=& - 4\delta \alpha - \frac{16}{\pi} \log\left(\frac{32}{27} \right)  \alpha u , \nonumber \\
	\beta(\tilde{\alpha}) &=& 4 \delta \tilde{\alpha} +  \frac{16}{\pi} \log\left(\frac{32}{27} \right) \tilde{\alpha} u \nonumber \\
	\beta(\delta) &=& -\frac{e^{4\gamma}}{4^{7} } J_{1} \left(  \alpha^{2} -   \tilde{\alpha}^{2}\right) - \frac{1}{2\pi^{2} } \left[\log\left(\frac{4}{3} \right) - \frac{5}{12} \right] u^{2} \nonumber. \\
\end{eqnarray}
Ignoring $\beta(\tilde{\alpha})$ the anisotropic sine-Gordon model was studied to one loop order by Grinstein\cite{Grinstein1981} and our results qualitatively agree with his. There are small differences in the numerical prefactors, but we suspect that this arises from differences in the renormalization procedure. It is not clear that the approximations used here amount to the same approximations used by Grinstein.\cite{Grinstein1981} As mentioned previously, $A=A(u,\vertex,\dvertex)$ is a renormalization group parameter and determined by the condition that $\beta(A)=0$.
\begin{eqnarray}
	 A(u,\alpha.\tilde{\alpha}) &&\simeq \nonumber \\
	&& -  \frac{2}{\pi^{2} } \left[ \log\left(\frac{4}{3} \right)-\frac{2}{12}  \right] u^{2} + \frac{e^{4\gamma} }{4^{5} } \frac{J_{2} }{32} \left( \alpha^{2}  -   \tilde{\alpha}^{2}  \right).\nonumber \\	
\end{eqnarray}
In particular the fix point of the theory is still the trivial fix point, $(\delta^{*},u^{*},\vertex^{*},\dvertex^{*} ) = (0,0,0,0)$. 

The one-loop RG equations, eq.\eqref{eq:one-loop-RG}, have the following structure. 
They have one critical line at $\alpha=\tilde \alpha=0$ and $u=0$. 
This is the six-vertex model line. Along this line the system is represented by the quantum Lifshitz model which has an exactly marginal operator, 
$(\nabla^2 \varphi)^2$  (and $\partial_t \varphi)^2$). 
Along this entire line the coupling constant $u \to 0$ very slowly (since $(\nabla \varphi)^4$ is a marginally irrelevant operator.) 
On the other hand, near the end of the six-vertex line the magnetic excitations with coupling constant $\tilde \alpha > 0$, are also marginally relevant, 
and this coupling constant grows under the RG. 
Thus, instead of a second fixed line, the phase boundary for $\tilde \alpha >0$ is either a first order transition (in which case there is no quantum criticality) 
or it flows to another quantum critical point which is not perturbatively accessible. Our one-loop RG cannot solve this problem. 
Since the $z=2$ dynamics is special and ``non-generic'' one possible solution is that the quantum critical theory is controlled by a $z=1$ fixed point 
(possibly with Ising criticality).

Treating $A$ as a renormalization group constant and fixing it to be $A=0$ for the $z=2$ quantum Lifshitz theory, it is easily seen that the renormalization group equations share many of the same properties as the $XY$-model. A critical surface is described by the equation, $f u^{2} =g (\alpha^{2}-\tilde{\alpha}^{2} )$. In the limit where there are no magnetic vortex excitations, the $\beta$-functions describe a line of fixed points and the charged excitations are marginally irrelevant operators. This is in agreement with earlier studies of the anisotropic sine-Gordon model.\cite{Grinstein1981} With the charged excitations suppressed, another line of fixed points is described by the $\beta$-functions and the magnetic vortices are marginally relevant operator that drive the system away from the critical point, much like in the generalized sine-Gordon model.\cite{Nienhuis1984,Boyanovsky1989} Like the generalized sine-Gordon model, this has two regimes depending on the sign of $\delta$. For $\delta\rightarrow -\infty$, the vertex operator is relevant and the vortex operator is irrelevant: $\alpha$ grows to large values while $\tilde{\alpha}$ tends to zero. In this case, the system is locked into the phase where for $\alpha<0$, $\varphi = (2n+1)\pi$ while for $\alpha>0$, $\varphi =0$. In the opposite limit, $\delta \rightarrow +\infty$, the opposite is true. Here the vertex operator is irrelevant and vortex operator is relevant. Now it is $\alpha$ that tends to zero while $\tilde{\alpha}$ grows large. Unfortunately, the argument is more complicated and the field $\Pi$ can not be thought of as being simply locked into a specific configuration. 

\subsection{RG flows and quantum stability analysis}

Finally, one can linearize the $\beta$-functions about the critical point. The renormalization group equations have the structure,
\begin{equation}
	\frac{\partial g_{i} }{\partial \tau} = - \sum_{j\leq k=1}^{4} C_{ijk} g_{j}g_{k},
\end{equation}
where $g_{1}=u,g_{2}=\alpha,g_{3}=\tilde{\alpha}$ and $g_{4}=\delta$ and $\tau=\log\vert \vec{x}\vert/a$ parameterizes the scale factor by which one changes the short distance cutoff. Now, suppose that one is close to the critical point and that $\alpha,\tilde{\alpha}\ll u$. Then at least for small $\tau$ near the critical surface, one can solve the renormalization group equations by eliminating terms $\mathcal{O}(\alpha^{2})$ and $\mathcal{O}(\tilde{\alpha}^{2})$. This leaves the set of two coupled differential equations,
\begin{eqnarray}
	\frac{\partial u}{\partial \tau} &=& - C_{111} u^{2} \nonumber \\
	\frac{\partial \delta}{\partial \tau} &=& - C_{411} u^{2},
\end{eqnarray}
which have the solution $u = u_{0}(1 + C_{111} u_{0} \tau)^{-1}$ and $\delta = \delta_{0} + C_{411} C_{111}^{-1} u$ where $u_{0},\delta_{0}$ are arbitrary integration constants. It is short work to solve the remaining differential equations,
\begin{eqnarray}
	\frac{\partial \alpha}{\partial \tau} &=& (- 4\delta - C_{212} u)\alpha \nonumber \\
	\frac{\partial \tilde{\alpha} }{\partial \tau} &=& (4\delta+C_{313} u)\tilde{\alpha} ,
\end{eqnarray}
from which one readily finds,
\begin{eqnarray}
	\alpha &=& \alpha_{0} (1+C_{111} u_{0}\tau)^{x} e^{-4\delta_{0} \tau} \nonumber \\
	\tilde{\alpha} &=& \tilde{\alpha}_{0} ( 1+ C_{111} u_{0} \tau)^{y} e^{4\delta_{0} \tau},
\end{eqnarray}
where $x = -(4 C_{411}+C_{212}C_{111})/C_{111}^{2} \simeq -1.96 $ and $y=-x$.

For the moment, we restrict ourselves to the region of parameter space where $\tilde{\alpha}=0$. With $\delta=0$, the fixed point $u=\delta=\alpha=0$ fixed point is reached as the scale $\tau\rightarrow \infty$. Specifically, as in agreement with Grinstein, \cite{Grinstein1981} we find that asymptotically, $u\sim A_{u}/\tau, \delta \sim A_{\delta}/\tau$ and $\alpha \sim A_{\alpha} \tau^{x}$ where $A_{u} \sim C_{111}^{-1}, A_{\delta} \sim C_{411}C_{111}^{-2}$ and $A_{\delta} \sim \alpha_{0} (C_{111}u_{0} )^{x}$. For $\delta <0$ one finds that $\alpha$ flows to infinity as the scale is changed. Flows stay close to the critical surface when $\tau$ is not too large. To be precise, for $\tau \sim 1/\delta_{0}$ the flows will remain near the critical surface as $\alpha$ does not flow to infinity too quickly. Recalling that $\tau = \log \vert \vec{x}\vert/a$ this gives an idea of how large the critical domains are in this phase and an idea of the correlation length. One finds that, $\xi/a \sim \exp\left[f(\delta)^{-\sigma}\right]$ where the critical exponent $\sigma=1$. On the surface where $\alpha=0$, similar considerations leads to identical conclusions only the regions $\delta>0$ and $\delta <0$ are exchanged.

\begin{figure}[hbt]
	\includegraphics[width=.4\textwidth]{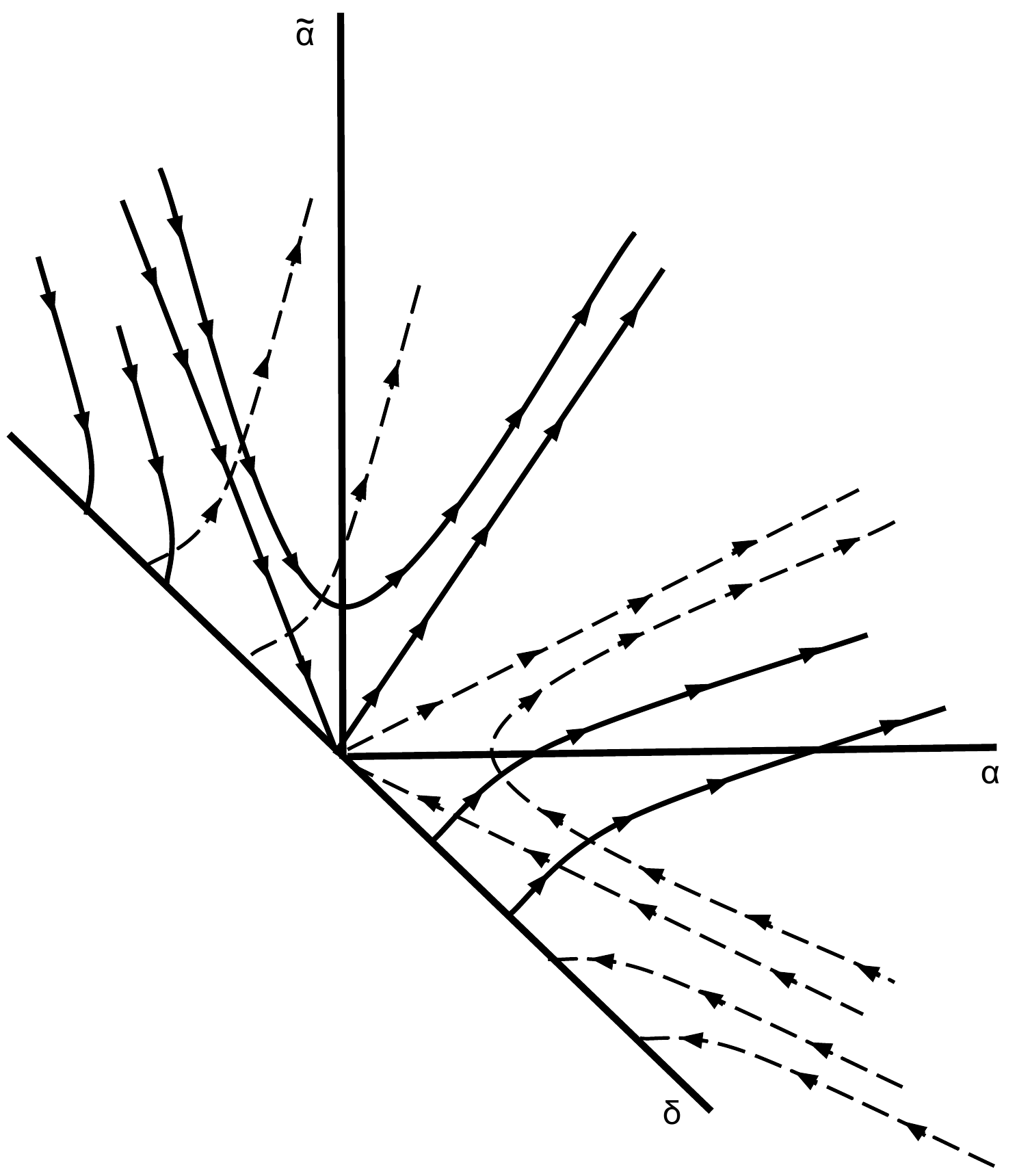}
	\caption{The renormalization group flows in the full quantum eight vertex model. The renormalization group flows are reminiscent of Kosterlitz-Thouless physics. Dashed lines are flow lines in the plane where $\tilde{\alpha}=0$ while the solid lines are in the plane with $\alpha=0$. There is a single fixed point at $(\delta,\alpha,\tilde{\alpha}) = (0,0,0)$. \label{fig:flow}  }
\end{figure}

Finally, we examine the situation where $\alpha,\tilde{\alpha}\neq0$. There is no critical point here since as $\tau\rightarrow \infty$ either $\alpha$ flows to larger values (when $\delta <0$) or its dual $\tilde{\alpha}$ flows to larger values (when $\delta >0$). However, again for values of $\tau$ that are modest, the flows stay near the critical surface. Focusing on the region $\delta >0$, one finds that for $\tau \sim \frac{1}{\delta_{0}}$,  $\tilde{\alpha}$ does not flow too quickly away. Once again, this gives a rough estimate to how the correlation length scales, namely, we find again that $\xi/a \sim \exp\left[ f(\delta)^{-\sigma}\right]$ with $\sigma=1$ again. Similar arguments may be applied to $\delta <0$. For more details, see Figure \ref{fig:flow}.

The main conclusion of the RG analysis is that the perturbed quantum multi-critical point of the quantum Lifshitz model is that when magnetic excitations are present they lead to a runaway behavior of the RG flows. Since at this point the theory has two marginally relevant perturbations the runaway process is very slow inducing logarithmic corrections-to-scaling in the correlators. The result of the runaway flow is that the quantum phase transition between the ordered phase and the $\mathbb{Z}_2$ topological phase is either a (fluctuation-induced) first order transition, or a continuous quantum phase transition controlled by a fixed point (presumably with `generic' $z=1$ dynamics) which is not accessible to the perturbative renormalization group theory that we have presented here. This result is in markedly different from the conclusions of Ardonne and coworkers,\cite{Ardonne2004} based on the equal-time behavior of the correlators in the Baxter wave function, and of the classical Monte Carlo simulations of Isakov and coworkers.\cite{Isakov2011} In the next section we analyze these issues.

\section{Dynamics and the RK condition}
\label{sec:disc}

In the previous section, we examined the parameter space of the $z=2$ quantum Lifshitz model in the presence of both magnetic and electric charge perturbations to one loop order. 
There we found that the magnetic perturbations at the quantum Baxter multicritical point lead to a runaway RG flow of the coupling constants which render the quantum phase transition either weakly (fluctuation-induced) first order or continuous but controlled by a non-perturbative fixed point. This result is markedly different than what happens in the classical Baxter model.\cite{Kadanoff-1979b,Nienhuis1984,Boyanovsky1989} We will see that the quantum dynamics is different deepening on whether the RK condition is preserved or not

Recently, there has been renewed interest in this model. 
Using classical simulations Isakov and coworkers found that the dynamics of the quantum Lifshitz model 
along the phase boundary between the ordered and topological phases of the model was in fact unstable and flowed from $z=2$ to 
values $z>2$ (albeit in a non-monotonic way).\cite{Isakov2011} The simulations along this phase boundary are classical in the sense that they follow the relaxational dynamics of the classical Baxter model. However, in the sense that we will now describe, these simulations nevertheless represent the quantum dynamics of the quantum eight vertex model of Ardonne {\it et al}.\cite{Ardonne2004}

The simulations of Ref.[\onlinecite{Isakov2011}] use the fact that Hamiltonians that obey the RK condition can also be used to represent classical stochastic processes of relaxation to thermal equilibrium.
For any discrete classical model, the classical dynamics can be described by a master equation.\cite{Glauber1963,Kawasaki-1966,Kadanoff-1968} If $p_{\alpha}$ is the probability of having a specific configuration of dimer 
(or arrows in the eight vertex model) $\alpha$, then the dynamics can be written down as a Master Equation
\begin{equation}
	\partial_t p_{\alpha}(t) = \sum_{\beta} W_{\alpha\beta} p_{\beta}(t)
	\label{eq:master}
\end{equation}
where the sum runs over configurations $\beta$ that differ from the configuration $\alpha$ by a single dimer flip. For the classical dimer model, the (positive) transition matrix $W_{\alpha\beta}$ is given by\cite{Henley1997,Henley2004}
\begin{equation}
	W_{\alpha\beta} = V_{\alpha} \delta_{\alpha\beta} - F_{\alpha\beta},
\end{equation}
where $V_{\alpha}$ counts the number of flippable dimers there are in a configuration $\alpha$ and $F_{\alpha\beta}$ dimer flip operator. 

The connection between the classical dynamics represented by the Master Equation, Eq.\eqref{eq:master} and a quantum Hamiltonian 
(acting on the same configuration space) is as follows.\cite{Siggia-1977} 
Let $p_\alpha(t)$ be the classical probability of a configuration $\alpha$ at time $t$, and let $w(\alpha)$ be a classical equilibrium 
Gibbs weights for the configurations $\alpha$. It then follows that, provided the transition matrix $W_{\alpha,\beta}$ obeys the detailed-balance condition, 

\begin{equation}
w_\alpha W_{\alpha,\beta}=W_{\alpha,\beta} w_\beta
\end{equation}
 (no sum over repeated labels is implied), 
then in the long time limit the system is guaranteed to reach the thermodynamic equilibrium state , 
\begin{equation}
\lim_{t \to + \infty} p_\alpha(t)=w_\alpha
\end{equation}
Now, upon a simple rescaling of the transition matrix, 
\begin{equation}
\tilde W_{\alpha,\beta}=w_\alpha^{-1/2} W_{\alpha,\beta} w_\beta^{1/2}
\end{equation}
and of the time-dependent probabilities, 
\begin{equation}
\tilde p_\alpha(t)=w_\alpha^{-1/2} p_\alpha(t)
\end{equation}
 allows us to identify the transformed 
Master Equation with a quantum mechanical evolution in imaginary time with a (hermitian) Hamiltonian 
\begin{equation}
H_{\alpha,\beta}=-\tilde W_{\alpha,\beta}
\end{equation}
and the wave functions become
\begin{equation}
\vert \Psi(t)\rangle=\sum_\alpha \tilde p_\alpha(t) \vert \alpha \rangle
\end{equation}
In particular the ground state wave function of this 
Hamiltonian has the form of a sum of local Gibbs weights, 
\begin{equation}
\vert \Psi_0\rangle=\sum_\alpha w_\alpha^{1/2} \vert \alpha \rangle
\end{equation}
By construction this state has exactly zero energy, $H \vert \Psi_0\rangle=0$, and the norm (squared) 
of the wave function is the classical equilibrium partition function,
\begin{equation}
||\Psi_0||^2=\sum_\alpha w_\alpha
\end{equation}
It is easy to see that the constraints required for the classical stochastic process to reach the equilibrium 
Gibbs state are equivalent to the RK condition that the Hamiltonian is a sum of projection operators. 
This mapping also implies that the spectrum of relaxation times of the classical evolution problem is the same as the spectrum of energy 
gaps in the quantum problems (with the gaps being the reciprocal of the relaxation times). 
This relation between classical stochastic processes and quantum Hamiltonians that satisfy the RK  condition has been extensively 
used to construct non-trivial 2D quantum states.\cite{Henley2004,Ardonne2004,Castelnovo-2006,Papanikolaou2007b}

Isakov and coworkers\cite{Isakov2011} used the relation between classical relaxation processes described by a 
Master Equation and 2D RK quantum Hamiltonians  to carry out extensive simulations. 
Here we will focus on their results for the quantum eight vertex (Baxter) model of Ref.[\onlinecite{Ardonne2004}] which obeys the RK condition. 

Along the six vertex model line, where the arrow configurations that violate the local conservation law are not allowed (hence the weight $d=0$)
 the classical simulations find that the dynamical exponent is $z=2$ as predicted by the quantum Lifshitz model.\cite{Ardonne2004} 
 This result also agrees with our findings in the preceding section in the absence of magnetic excitations, which reproduce the earlier 
 RG results in the same regime.\cite{Grinstein1981,Vishwanath-2004,Fradkin2004}

On the other hand, along the phase boundary between the ordered phase and the $\mathbb{Z}_2$ topological phase, parametrized by $c^2=d^2+2$, 
their classical simulation obtains a 
dynamical exponent $z>2$. This result is consistent with the fact that the classical dynamics along the six vertex model line, 
which has a local conservation law and consequently the quantum Hamiltonian (and hence the Liouville operator) has a continuous symmetry. 
In this case the classical critical dynamics  
is in the  model B class in the Hohenberg-Halperin classification,\cite{Hohenberg-Halperin} and hence has $z=2$ along this entire line. 
In contrast, along the 
phase boundary between the $\mathbb{Z}_2$ topological phase and the the ordered phase the local conservation law is broken but the quantum 
Hamiltonian has a local $\mathbb{Z}_2$ symmetry. Thus in this case the universality class of the classical  critical dynamics is expected to be model A. 
Their result of $z>2$ is consistent with the inequalities that the dynamical exponent is known to obey in the case of 
Model A dynamics.\cite{Hohenberg-Halperin,Katzav-2011} Moreover at a special point on this phase boundary, $(c^2,d^2)=(\sqrt{2}+1,\sqrt{2}-1)$, the norm 
squared of the ground state wave function is equal to the partition function of two decoupled 2D classical critical Ising models. 
The value of the dynamical critical exponent obtained by Isakov and coworkers turns out to be $z=2.196$ which is consistent with the best estimates for 
$z$ in the 2D critical Ising model with model A dynamics.
However what is surprising is the fact that they find an exponent $z$ which varies continuously (and non-monotonically) along this entire phase boundary, 
rising from a value close to $2$ at $(c^2,d^2)=(2,0)$ (the KT transition of the six vertex model), peaking at the decoupled Ising point mentioned above 
and slowly falling off past that point to values still obeying $z\geq 2$.

Isakov and coworkers also considered the effects of perturbations to the eight vertex model Hamiltonian (at the Ising decoupling point) that violate the RK condition. 
In this case they did quantum Monte carlo simulations and found that for all values of this RK-violating perturbation the dynamics obeys $z=1$, instead of 
$z=2.196$ when the RK condition holds. The value $z=1$ is consistent with the critical behavior of the 2D {\em quantum} Ising model (in a transverse field), 
whose quantum Hamiltonian is (essentially) the same a transfer matrix of the 3D classical Ising model. The quantum Ising model is well known to be the dual 
of the $2+1$-dimensional $\mathbb{Z}_2$ (Ising) gauge theory. This is also consistent with the fact that the Kitaev Toric Code point is equivalent to the 
ultra-deconfined limit of the $\mathbb{Z}_2$ gauge theory. This result suggests that perturbations to the  2D quantum Hamiltonian that violate the RK 
condition lead to a system with `generic' $z=1$ `relativistic' dynamics. 

This result of Isakov and coworkers is consistent with the renormalization group analysis presented in the preceding section. There we showed  that by
perturbing the quantum Lifshitz model in the vicinity of 
the KT transition, where both the electron and magnetic perturbations are marginally relevant and lead to a runaway RG flow. Although the quantum Lifshitz model 
itself is consistent with the RK condition, generic perturbations are not consistent with this constraint ,and in general kill the quantum criticality, although it is still 
possible to reach this multicritical point by fine-tuning of the parameters. However, in the presence of magnetic perturbations this is no longer possible as their 
effective coupling constants flow to strong coupling. Thus, in this case the physics is controlled by a fixed point which not accessible in perturbation theory.  

Although the quantum Monte Carlo simulations of Isakov and coworkers were done only near the special decoupled Ising point, combined with our 
RG analysis they suggest that in general, if the
 quantum phase transition between the $\mathbb{Z}_2$ topological phase and the ordered phase remains continuous, then it is always in the universality 
 class of the deconfinement quantum phase transition of the $2+1$-dimensional Ising gauge theory (which has $z=1$). Contrary to the case of the quantum Lifshitz 
 model, the quantum phase transition of the $\mathbb{Z}_2$ gauge theory (or, equivalently the 3D classical Ising model) does not have any marginal operators and 
 hence the critical exponents are fixed and no longer can vary as functions of any parameters.
 
 What remains to be understood is the result of Isakov and coworkers that at critical dynamics of the quantum Baxter model which satisfies the RK condition has a 
 dynamical exponent $z$ that can vary smoothly along the phase boundary between the topological and the ordered phases. This result is puzzling for the following reasons. Let us consider first what happens along the six vertex model line where the simulations obtain a dynamical exponent $z=2$, which is consistent with the quantum Lifshitz scaling. The classical critical dynamics of the coarse-grained height field $\varphi(\vec x, t)$ can be expressed in terms of a Langevin equation of the form\cite{Hohenberg-Halperin}
 \begin{equation}
\partial_t{\varphi}(\vec{x},t) = -\frac{\Omega }{2}  \frac{\delta S[\varphi]}{\delta \varphi}+ \xi(\vec{x},t)
\label{eq:langevin}
\end{equation}
where the classical equilibrium state has a Gibbs distribution with weight $\exp(-S[\varphi])$, and
 is a Gaussian distributed noise with the properties,
\begin{eqnarray}
	\langle \xi(\vec{x},t) \rangle &=& 0, \nonumber \\ 
	 \langle \xi (\vec{x},t) \xi(\vec{x}',t') \rangle &=& \Omega \delta^{2}(\vec{x}-\vec{x}') \delta(t-t'),
\end{eqnarray}
where $\Omega$ is the standard deviation of the noise. Along the six-vertex line we can take $S_{6v}[\varphi]$ to have the (Euclidean) sine-Gordon form
\begin{equation}
S_{6v}[\varphi]=\int d^2x \left( \frac{K}{2} (\nabla \varphi)^2+ g_n \cos n \varphi\right)
\label{eq:S6v}
\end{equation}
Here $g_n$ is the coupling constant for electric charge excitations, labelled by $n$, in the classical problem.

It is possible to express this stochastic process in terms of a path integral in $d=2$ space dimensions plus time (``Monte Carlo time'').\cite{MSR} The condition that the process converges to the equilibrium distribution and satisfies the Langevin equation leads to a path-integral with a local supersymmetric action.\cite{DeDominicis1978,ZinnJustin} In particular, it is the supersymmetry that guarantees that the process relaxes to the chosen equilibrium state. Thus, the RK condition for the quantum Hamiltonian, that requires the ground state to have exactly zero energy,  implies a form of supersymmetry. On the other hand perturbations to the quantum Hamiltonian that break the RK condition also break the supersymmetry and, hence, are not consistent with a Langevin relaxation processes.  Hence generic perturbations to the RK Hamiltonian cannot be simulated by classical relaxation processes.

One important property of the Langevin process of Eq.\eqref{eq:langevin} for the equilibrium state defined by Eq.\eqref{eq:S6v}, is that it is {\em local}. In other words, electric charge perturbations are local. We know from the simulations of Isakov and coworkers that in this case the dynamics has $z=2$, which is the same as in the quantum Lifshitz model. 
One the other hand if we want the equilibrium state of the classical baxter model we need to involve both electric charge and magnetic charge excitations.\cite{Kadanoff-1979b} The classical (Euclidean) action now has the form (for suitable choices of $n$ and $m$)
\begin{equation}
S_{8v}[\varphi]=\int d^2x \left(\frac{K}{2} (\nabla \varphi)^2+g_n \cos n \varphi+\tilde g_m \cos m \tilde \varphi \right)
\label{eq:S8v}
\end{equation}
where $m$ are the magnetic charges. The field $\tilde \varphi$ is the dual field and it is related to the coarse-grained field $\varphi$ by the Cauchy-Riemann relation
\begin{equation}
\partial_\mu \tilde \varphi=\epsilon_{\mu \nu} \partial_\nu \varphi
\end{equation}
where $\epsilon_{\mu \nu}$ is the second rank Levi-Civita tensor. 

The solution of the Cauchy-Riemann relation implies that the fields $\varphi$ and its dual $\tilde \varphi$ are non-local with respect to each other. This means that the Langevin process treats both types of excitations in a very different way and, in particular, the presence of magnetic excitations makes the process non-local. Thus, while the equilibrium partition function is invariant under duality (which exchanges electric and magnetic charges) the dynamics is not. 
It is reasonable to speculate that 
this difference and the inherent non-locality in the dynamics could possibly lead to a value of $z>2$. Nevertheless, this still does not explain in any obvious way why isakov and coworkers  have a value of $z$ that varies continuously along the phase boundary of the eight vertex model.

\section{Conclusions}
\label{sec:conclusions}

Motivated by the work by Isakov et. al.\cite{Isakov2011} and their surprising observation of a continuing varying dynamical exponent in the quantum Lifshitz theory, we have performed a perturbative RG analysis of the $2+1$-dimensional quantum field theory. We added both electric perturbations and magnetic perturbations what are both marginal at the same point along the six vertex line of the quantum Lifshitz model. In order to do the perturbative RG analysis, we have extended the operator product expansion to the case of anisotropic critical points.  In the case where magnetic vortices are ignored, our results are in agreement with those found by Grinstein.\cite{Grinstein1981} However, some minor differences that were found presumably  arise from differences in how the renormalization scheme was implemented.

By an analysis of the RG equations we saw that the perturbed quantum Lifshitz theory has a single fixed line and that the dynamical critical exponents is $z=2$. 
On the other hand, the magnetic vortices were marginally relevant and hence the theory has potentially two marginally relevant operators. In such situations, two things may happen: (1) there is no quantum criticality and the quantum phase transition become first order, or (2) the system flows to a new fixed point, which is not perturbatively accessible. We guess that this may be the $z=1$ theory, but our perturbative RG calculation at one lop can not be used to access this theory. Even given these caveats, our results do not seem to be compatible with a continuously varying dynamical exponent as implied by the numerical simulations carried out by Isakov and coworkers.\cite{Isakov2011} It remains to be understood how the how the continuously varying dynamical exponent found by Isakov and coworkers can be consistent with a field theoretical perspective.

\begin{acknowledgments}
We thank Paul Fendley and Shivaji Sondhi for illuminating discussions. This work was supported in part by NSF grant PHY-1005429 (BH) and by the National Science Foundation under the grants DMR 0758462 and  DMR-1064319 at the University of Illinois (EF). \end{acknowledgments}

\appendix
\begin{widetext}
\section{Useful Quantities}
\label{append:useful_quants}

The Green function is given by
\begin{equation}
 	G(\vec{r},t) = \langle \varphi(\vec{r},t)\varphi(0) \rangle = -\frac{1}{8\pi \kappa} \left[ \log\left( \frac{\vert \vec{r} \vert^{2}}{a^{2} } \right) + \Gamma\left(0,\frac{\vert \vec{r} \vert^{2} }{4\kappa \vert t \vert} \right) \right]
\end{equation}
Defining $s = \vert \vec{r}\vert^{2} /\kappa \vert t\vert $ and $f(s) = (1-e^{-s/4})$,
\begin{eqnarray}
	\partial_{i} G(\vec{r},t) &=& - \frac{ r_{i} }{4\pi \kappa r^{2} } f(s) \nonumber \\
	\partial_{i}\partial_{j} G(\vec{r},t) &=& -\frac{1}{4\pi\kappa r^{2} } \left[ \delta_{ij} f(s) - \frac{2 r_{i}r_{j} }{r^{2} } \left( f(s) - \frac{s}{4}e^{-s/4} \right) \right] \nonumber \\
	\nabla^{2} G(\vec{r},t) &=&  -\frac{1}{8 \pi \kappa r^{2} } s e^{-s/4} \\ 
	\partial_{i} \nabla^{2} G(\vec{r},t) &=& \frac{ r_{i}}{16\pi \kappa r^{4} } s^{2} e^{-s/4} \nonumber \\
	\partial_{i}\partial_{j} \nabla^{2} G(\vec{r},t) &=& \frac{1}{16 \pi \kappa r^{4} } s^{2} e^{-s/4} \left[ \delta_{ij} - \frac{r_{i}r_{j} }{2 r^{2} } s\right] \nonumber \\
	\nabla^{4}G(\vec{r},t) &=& \frac{1}{8\pi \kappa r^{4} } s^{2} e^{-s/4} \left[ 1-\frac{s}{4} \right] \nonumber 
\end{eqnarray}

\section{Computation of the fusion coefficients}
\label{append:OPE}
In this appendix, we give a detailed account of the computation of the non-zero fusion coefficients in the OPE. Defining $\vec{x}-\vec{y} = 2\vec{r}$ and $\vec{x}+\vec{y} =2\vec{R}$ and $t_{x}-t_{y}=2t_{r}$ and $t_{x}+t_{y}=2t_{R}$ recall that we will be interested in the limit where $R\gg r$ and $t_{R}\gg t_{r}$. We are interested in the short distant ($r\rightarrow 0, t_{r}\rightarrow 0$ with $s=r^{2}/\ka \vert t\vert$ fixed) divergent behavior of the three point functions. Useful quantities used in this section are defined as, $R_{\pm}=(\vec{R}-\vec{z}) \pm \vec{r}$, $T_{\pm}=(t_{R} - t_{z} ) \pm t_{r}$ and $s_{\pm} = \vert R_{\pm}\vert^{2}/\kappa \vert T_{\pm} \vert$.

 \subsection{Fusion of $\mathcal{O}_{q} \mathcal{O}_{-q}$ }

In this section, we provide details of the OPE for two vertex operators $\langle \mathcal{O}_{1}\mathcal{O}_{-1}\rangle $. These correlation functions can be computed in much the same spirit as the two dimensional, critical XY model at the KT transition. Here, we look at the fusion of two vertex operators into (a) the $(\nabla\varphi)^{2} $ operator, (b) the $(\nabla^{2}\varphi)^{2}$ operator and (c) the $(\nabla\varphi)^{4}$ operator. 

\subsubsection{A. $(\nabla\varphi)^{2}$} 

We point split the operator $(\nabla\varphi)^{2}(z)=\lim_{z_{i}\rightarrow z} \nabla_{z_{1} }\varphi(z_{1}) \nabla_{z_{2}} \varphi(z_{2}) $. It can be shown that upon subtracting self energy contributions that the only non-vanishing contribution is given by,
\begin{eqnarray*}
	 \langle  \mathcal{O}_{1}(\vec{x},t_{x})\mathcal{O}_{-1}(\vec{y},t_{y}) \, : (\nabla\varphi)^{2}(z): \rangle \simeq -\frac{1 }{2^{2} } e^{-q^{2} G_{reg}(x-y)} \lim_{\vec{z}_{i}\rightarrow \vec{z} }\partial_{i}^{(z_{1}) }\partial^{i,\,(z_{2}) } \prod_{i=1}^{2} \left[ G(x-z_{i})-G(y-z_{i})\right] .
\end{eqnarray*}
 With the aforementioned variables $R_{\pm},T_{\pm}$ and $s_{\pm}$, it is easily seen using the results from Appendix \ref{append:useful_quants} that,
\begin{eqnarray*}
	\left(\partial_{i} ( G(x-z_{i}) - G(y-z_{i}))  \right)^{2}  =  \frac{1}{(4\pi\ka)^{2} } \left( \frac{f^{2} (s_{+}) }{R^{2}_{+} } + \frac{f^{2} (s_{-}) }{R^{2}_{-} }  - \frac{2R_{+}^{i} R_{i, -}}{R_{+}^{2}R_{-}^{2} } f(s_{+}) f(s_{-}) \right).
\end{eqnarray*}
 In the limit where $R\gg r, t_{R}\gg t_{r}$ we have the simplification $s_{+}\simeq s_{-} = s_{R-z}$ and $R_{+}\simeq R_{-} = R-z$. Thus arriving at the result,
 \begin{equation}
	 \langle  \mathcal{O}_{1}(\vec{x},t_{x})\mathcal{O}_{-1}(\vec{y},t_{y}) \, : (\nabla\varphi)^{2}(z): \rangle \simeq -\frac{ r^{2}    e^{-G_{reg}(r)} }{(4\pi\ka)^{2} }\frac{f^{2}(s_{R-z} )}{(R-z)^{4} }.
\end{equation}

\subsubsection{Fusion into $(\nabla^{2}\varphi)^{2}$ }

Once again point splitting the operator $(\nabla^{2}\varphi)^{2}(z)  =\lim_{z_{i}\rightarrow z} \nabla^{2}_{z_{1} }\varphi(z_{1}) \nabla^{2}_{z_{2}} \varphi(z_{2})$ and subtracting divergent self-energy terms, one arrives at,
 \begin{eqnarray*}
 	 \langle  \mathcal{O}_{1}(\vec{x},t_{x})\mathcal{O}_{-1}(\vec{y},t_{y}) \, : (\nabla^{2}\varphi)^{2}(z): \rangle \simeq -\frac{1 }{2^{2} } e^{-q^{2} G_{reg}(r)} \nabla^{2}_{z_{1} } \nabla^{2}_{z_{2}}  \prod_{i=1}^{2} \left[ G(x-z_{i})-G(y-z_{i})\right] .
 \end{eqnarray*} 
 The non-trivial quantity to examine is then (in the limit $R\gg r$ and $t_{R}\gg t_{r}$),
 \begin{eqnarray*}
	\nabla^{2}_{z} \left[G(x-z) - G(y-z) \right] \simeq \frac{e^{-s_{R-z}/4}}{(8\pi\ka)} \left( \frac{1}{\vert T_{+}\vert } - \frac{1}{\vert T_{-}\vert  } \right).
 \end{eqnarray*}
The difference in $\vert T_{+}\vert^{-1}$ and $\vert T_{-}\vert^{-1}$ can be studied by writing $\vert T_{+} \vert = \sqrt{(t_{R}-t_{z}+t_{r})^{2}} $ and  $\vert T_{-} \vert = \sqrt{(t_{R}- t_{z}-t_{r})^{2}} $. Hence, we arrive at the result,
 \begin{equation}
 	 \langle  \mathcal{O}_{1}(\vec{x},t_{x})\mathcal{O}_{-1}(\vec{y},t_{y}) \, : (\nabla^{2}\varphi)^{2}(z): \rangle \simeq -\frac{ e^{-q^{2} G_{reg}(r) }  t^{2}_{r} }{ (8\pi\ka)^{2} }  \frac{e^{-s_{R-z}/2 }}{(t_{R}-t_{z})^{4} }.
\end{equation}

\subsubsection{Fusion into $(\nabla\varphi)^{4}$ }

The computation of this fusion coefficient is done as the previous two cases. We point split the operator and subtracting the self-energy terms, its possible to show that the non-vanishing contribution is given by,
\begin{eqnarray*}
	 \langle  \mathcal{O}_{1}(\vec{x},t_{x})\mathcal{O}_{-1}(\vec{y},t_{y}) \, : (\nabla\varphi)^{4}(z): \rangle = \frac{q^{4} }{2^{4}}e^{-q^{2} G_{reg}(x-y)} \lim_{\vec{z}_{i}\rightarrow \vec{z} }\partial_{i}^{(z_{1}) }\partial^{i,\,(z_{2}) } \partial_{j}^{(z_{3}) } \partial^{j,\,(z_{4}) } \prod_{i=1}^{4} \left[ G(x-z_{i})-G(y-z_{i})\right] .
 \end{eqnarray*} 
By a similar set of manipulations, in the limit $R\gg r$ and $t_{R}\gg t_{r}$, its possible to show that
 \begin{equation}
	\langle  \mathcal{O}_{1}(\vec{x},t_{x})\mathcal{O}_{-1}(\vec{y},t_{y}) \, : (\nabla\varphi)^{4}(z): \rangle \simeq \frac{q^{4} e^{-q^{2} G_{reg}(r) }  r^{4}  }{(4\pi\ka)^{4} }   \frac{f^{4}(s_{R-z})}{(R-z)^{8} } .
\end{equation}


\subsection{Fusion of $:(\nabla\varphi)^{4}: :(\nabla\varphi)^{4}: $}
In this section, details are provided for the fusion of two operators $\langle :(\nabla\varphi)^{4}: :(\nabla\varphi)^{4}:\rangle$. The non-zero fusion channels are (a) the $(\nabla\varphi)^{2}$ operators, (b) the $(\nabla^{2}\varphi)^{2}$ operator and finally (c) the $(\nabla\varphi)^{4}$ operator. Useful quantities defined in this appendix are  $D_{i}^{k} = \left( \partial_{i}^{(x_{1})} \partial^{k, (y_{1}) } G(x_{1}-y_{1})  \right)$ and $d^{j} =\left( \partial^{j, (x_{4})} \nabla^{2}_{(z_{2}) } G(x_{4}-z_{2})  \right)$.

\subsubsection{Fusion into $(\nabla\varphi)^{2}$}

Point splitting each of the operators present, this channel can be computed as,
\begin{eqnarray*}
	\lim_{\vec{x}_{i}\rightarrow \vec{x} } \lim_{\vec{y}_{i}\rightarrow \vec{y} } \lim _{\vec{z}_{i}\rightarrow\vec{z}}\left( \partial_{i}^{(x_{1})} \partial^{i, (x_{2})} \partial_{j}^{(x_{3})} \partial^{j, (x_{4})} \right)\left( \partial_{k}^{(y_{1})} \partial^{k, (y_{2})} \partial_{\ell}^{(y_{3})} \partial^{\ell, (y_{4})} \right) \left( \partial_{m}^{(z_{1})} \partial^{m, (z_{2})} \right)  \nonumber \\
	\times \langle \varphi(x_{1})\varphi(x_{2})\varphi(x_{3})\varphi(x_{4}) \varphi(y_{1})\varphi(y_{2})\varphi(y_{3})\varphi(y_{4}) \varphi(z_{1})\varphi(z_{2}) \rangle 
\end{eqnarray*}
Simple combinatorics reveals that there are two classes of possible contractions for the matrices where indices are contracted as $\textrm{Tr}(\mathbf{D D} ) \textrm{Tr}(\mathbf{DDD})$ and one where the indices are all contracted together $\textrm{Tr}(\mathbf{DDDDD})$.
The result follows.
 \begin{equation}
 	\langle :(\nabla\varphi)^{4}(\vec{x}):  :(\nabla\varphi)^{4}(\vec{y}):  :(\nabla\varphi)^{2}(\vec{z}):\rangle \simeq  -\frac{4 }{(\pi\ka)^{2} r^{3} }\left[ \frac{2}{12} - \log\left(\frac{4}{3} \right) \right]    \, \, \langle :(\nabla\varphi)^{2}(\vec{R})::(\nabla\varphi)^{2}(\vec{z}): \rangle
\end{equation}

\subsubsection{Fusion into $(\nabla^{2}\varphi)^{2}$}

As in the previous two cases, one needs to compute the quantity,
\begin{eqnarray*}
	\lim_{\vec{x}_{i}\rightarrow \vec{x} } \lim_{\vec{y}_{i}\rightarrow \vec{y} } \lim _{\vec{z}_{i}\rightarrow\vec{z}} \left( \partial_{i}^{(x_{1})} \partial^{i, (x_{2})} \partial_{j}^{(x_{3})} \partial^{j, (x_{4})} \right)\left( \partial_{k}^{(y_{1})} \partial^{k, (y_{2})} \partial_{\ell}^{(y_{3})} \partial^{\ell, (y_{4})} \right) \left( \nabla^{2}_{(z_{1})} \nabla^{2}_{(z_{2})} \right) \nonumber \\
	\times \langle \varphi(x_{1})\varphi(x_{2})\varphi(x_{3})\varphi(x_{4}) \varphi(y_{1})\varphi(y_{2})\varphi(y_{3})\varphi(y_{4}) \varphi(z_{1})\varphi(z_{2}) \rangle 
\end{eqnarray*}
Again, one has to treat the different contraction of indices. One representative from this set is,
\begin{eqnarray*}
	\left( \partial_{i}^{(x_{1})} \partial^{k, (y_{1}) } G(x_{1}-y_{1})  \right)\left( \partial^{i, (x_{2})} \partial^{(y_{2}) }_{k} G(x_{2}-y_{2})  \right) \left( \partial_{j}^{(x_{3})} \partial^{\ell, (y_{3}) } G(x_{3}-y_{3})  \right) \nonumber \\
	\times\left( \partial^{j, (x_{4})} \nabla^{2}_{(z_{1}) } G(x_{4}-z_{2})  \right) \left( \partial_{\ell}^{(y_{4}) } \nabla^{2}_ {(z_{2}) } G(y_{3}-z_{3})  \right) 
\end{eqnarray*}
There are of course many diagrams that yield the same contraction of indices. These are similar to the previous case. One finds that the possible contractions are given by $\textrm{Tr }(\mathbf{D}(r)\cdot \mathbf{D}(r)) \times \vec{d}(R_{+})\cdot \mathbf{D}(r) \cdot \vec{d}(R_{-} )$ and by $\vec{d}(R_{+}) \cdot \mathbf{D}\cdot \mathbf{D}\cdot \mathbf{D} \cdot \vec{d}(R_{-}) $. Hence one finds that there is a non-vanishing contribution from this correlation function to this fusion channel.
\begin{equation}
	\langle :(\nabla\varphi)^{4}(\vec{x}):  :(\nabla\varphi)^{4}(\vec{y}):  :(\nabla^{2}\varphi)^{2}(\vec{z}):\rangle \simeq\frac{1}{(\pi\ka)^{2} r }\left[ - \frac{5}{12} +  \log \left(\frac{4}{3} \right) \right]  \,\, \langle :(\nabla^{2}\varphi)^{2}(\vec{R}): :(\nabla^{2}\varphi)^{2}(\vec{z}): \rangle.
\end{equation}

\subsubsection{Fusion into $(\nabla\varphi)^{4}$}

To compute this correlation function, one needs to examine,
\begin{eqnarray*}
	\lim_{\vec{x}_{i}\rightarrow \vec{x} } \lim_{\vec{y}_{i}\rightarrow \vec{y} } \lim _{\vec{z}_{i}\rightarrow\vec{z}}  \left( \partial_{i}^{(x_{1})} \partial^{i, (x_{2})} \partial_{j}^{(x_{3})} \partial^{j, (x_{4})} \right)\left( \partial_{k}^{(y_{1})} \partial^{k, (y_{2})} \partial_{\ell}^{(y_{3})} \partial^{\ell, (y_{4})} \right) \left( \partial_{m}^{(z_{1})} \partial^{m, (z_{2})} \partial_{n}^{(z_{3})} \partial^{n, (z_{4})} \right) \nonumber \\
	\times \langle \varphi(x_{1})\varphi(x_{2})\varphi(x_{3})\varphi(x_{4}) \varphi(y_{1})\varphi(y_{2})\varphi(y_{3})\varphi(y_{4}) \varphi(z_{1})\varphi(z_{2})\varphi(z_{3})\varphi(z_{4}) \rangle 
\end{eqnarray*}
One has to treat the different contraction of indices carefully. The combinatorics for involved is more complicated than the other two, but careful counting reveals that there are four channels. These are simply given by the different ways of grouping a contraction of six matrices: $\left[ \textrm{Tr} (\mathbf{DD})\right]^{3}, \textrm{Tr} (\mathbf{DDDD}) \textrm{Tr}(\mathbf{DD}), \textrm{Tr} (\mathbf{DDDDDD })$ and $\textrm{Tr} (\mathbf{DDD}) \textrm{Tr} (\mathbf{DDD})$. In the limit $R\gg r$, we find a non-vanishing fusion coefficient,
\begin{equation}
	\langle :(\nabla\varphi)^{4}(\vec{x}):  :(\nabla\varphi)^{4}(\vec{y}):  :(\nabla\varphi)^{4}(\vec{z}): \rangle \simeq \frac{41}{16 \pi\ka} \frac{1}{r }   \,\, \langle  :(\nabla\varphi)^{4}(\vec{R}):  :(\nabla\varphi)^{4}(\vec{z}): \rangle .
\end{equation}


\subsection{Fusion of $\tilde{\mathcal{O}}_{8} \tilde{\mathcal{O}}_{-8} $ }

In this section, we compute the fusion of two dislocation operators $\tilde{\mathcal{O}}_{8}\tilde{\mathcal{O}}_{-8} $. Non-zero fusion channels are in the (a) $(\nabla\varphi)^{2}$ channel, (b) the $(\nabla^{2}\varphi)^{2}$ and the (c) $(\nabla\varphi)^{4}$ channel. 

\subsubsection{Fusion into $(\nabla\varphi)^{2}$}

Point splitting the operator $(\nabla\varphi)^{2}$, this quantity can be computed by taking functional derivatives of the quantity,
\begin{eqnarray*}
	\lim_{z_{i}\rightarrow z} \partial_{i, (z_{1})} \partial^{i}_{ (z_{2}) } \langle \tilde{\mathcal{O} }_{8}(\vec{x}) \tilde{\mathcal{O} }_{-8}(\vec{y})  :\varphi(\vec{z}_{1}) \varphi(\vec{z}_{2} ) \rangle &=& \frac{1}{Z_{0} } \int D\varphi D\Pi \, \, \textrm{exp}\Bigg( \int d^{3}\vec{x}'\, \frac{1}{2} \Pi \dot{\varphi} - \frac{1}{2} \Pi^{2} - \frac{\kappa^{2} }{2} (\nabla^{2}\varphi)^{2}  \\
	&& + A(\vec{x}',t) \Pi(\vec{x}',t)  + J(\vec{x}',t')\varphi(\vec{x}',t') \Bigg) \nonumber 
\end{eqnarray*}
where
\begin{eqnarray}
	A(\vec{x}',t') = 2 im\left( \textrm{arg}(\vec{x'}-\vec{x} ) \delta(t'-t_{x} ) - \textrm{arg}(\vec{x'}-\vec{y}) \delta(t'-t_{y}) \right)
\end{eqnarray}
Integrating out $\Pi$ and then, redefining $\varphi = \varphi' + \alpha(\vec{x}',t')$ where
\begin{eqnarray}
	\alpha(\vec{x}',t') =2  i m \left( \textrm{arg}(\vec{x'}-\vec{x} ) \Theta(t'-t_{x} ) - \textrm{arg}(\vec{x'}-\vec{y}) \Theta(t'-t_{y}) \right), 
\end{eqnarray}
one is able to eliminate the field $A(\vec{x'},t')$ at the expense of adding the field $\alpha(\vec{x}',t')$. 
\begin{eqnarray*}
	&& \lim_{z_{i}\rightarrow z} \partial_{i, (z_{1})} \partial^{i}_{ (z_{2}) } \langle \tilde{\mathcal{O} }_{8}(\vec{x}) \tilde{\mathcal{O} }_{-8}(\vec{y})  :\varphi(\vec{z}_{1}) \varphi(\vec{z}_{2} ) \rangle \\
	&&= \frac{e^{- 2\pi m^{2} \kappa^{2} G_{reg}(x-y)} }{Z_{0}} \exp\Bigg( \int d^{3}\vec{x}'d^{3}\vec{y'}\, \frac{1}{2} J(x')G(x'-y') J(y') + \kappa^{2} \nabla^{4} \alpha(x')G(x'-y') J(y') - J(x')\alpha(x') \Bigg)
\end{eqnarray*}
Taking functional derivatives, there are many terms, however  the only terms that survive are those were the points $\vec{z}_{i}$ are not connected to each other. Namely, we are left with,
\begin{eqnarray*}
	 \lim_{z_{i}\rightarrow z} \partial_{i,(z_{1})} \partial^{i}_{(z_{2})} \langle \tilde{\mathcal{O} }_{8}(\vec{x}) \tilde{\mathcal{O} }_{-8}(\vec{y})  :\varphi(\vec{z}_{1}) \varphi(\vec{z}_{2} ): \rangle &=& e^{- 2\pi m^{2} \kappa^{2} G_{reg}(x-y)} \\
	&&  \quad  \lim_{z_{i}\rightarrow z} \partial_{i, (z_{1})} \partial^{i}_{(z_{2})} \Bigg[ \kappa^{2} \int d^{3}\vec{x}' \, \, \nabla^{4} \alpha(x') G(x'-z_{1}) -\alpha(z_{1} ) \Bigg] \\
	 &&\quad \times \Bigg[ \kappa^{2} \int d^{3}\vec{x}'' \,\, \nabla^{4} \alpha(x'') G(x'' -z_{2} ) - \alpha(z_{2} ) \Bigg] .
\end{eqnarray*}
In the limit $r\ll R$ and to \emph{leading} divergent behavior in $R$, one finds
\begin{equation}
	\langle \tilde{\mathcal{O} }_{8}(\vec{x}) \tilde{\mathcal{O} }_{-8}(\vec{y})  :(\nabla\varphi)^{2}(\vec{z}): \rangle = -4 \ka^{4} m^{2} e^{- 2\pi m^{2} \kappa^{2} G_{reg}(r)} \left[ \frac{1}{(4\pi \ka)^{2}} 4r^{2} \frac{ f^{2}(s_{R})}{R^{4} } + \mathcal{O}(1/R^{3} ) + \mathcal{O}(1/R^{2}) \right]
\end{equation}

\subsubsection{Fusion into $(\nabla^{2}\varphi)^{2}$}
In a very similar set of manipulations, the fusion of two dislocation operators into the $(\nabla^{2}\varphi)^{2}$ operator can be computed. One finds the relevant quantity is given by, 
 \begin{eqnarray*}
	 \lim_{z_{i}\rightarrow z} \nabla^{2}_{(z_{1})} \nabla^{2}_{(z_{2})} \langle \tilde{\mathcal{O} }_{8}(\vec{x}) \tilde{\mathcal{O} }_{-8}(\vec{y})  :\varphi(\vec{z}_{1}) \varphi(\vec{z}_{2} ): \rangle &\simeq& e^{- 2\pi m^{2} \kappa^{2} G_{reg}(x-y)} \\
	&&  \quad  \lim_{z_{i}\rightarrow z} \nabla^{2}_{(z_{1})} \nabla^{2}_{(z_{2})} \Bigg[ \kappa^{2} \int d^{3}\vec{x}' \, \, \nabla^{4} \alpha(x') G(x'-z_{1}) -\alpha(z_{1} ) \Bigg] \\
	 &&\quad \times \Bigg[ \kappa^{2} \int d^{3}\vec{x}'' \,\, \nabla^{4} \alpha(x'') G(x'' -z_{2} ) - \alpha(z_{2} ) \Bigg].
\end{eqnarray*}
Now, I note that $\nabla^{2}_{(z_{i} ) } \alpha(z_{i}) = 2\pi\left(  \delta^{(2)} (z_{i}-x) \Theta(\tau_{i} -t_{x} ) -  \delta^{(2)}(z_{i}-y) \Theta(\tau_{i} -t_{y} )\right)$ which is vanishing when $\vec{r} \ll \vec{R}$. Hence, in this case, the only terms that survive come from when the Laplacian hits $G(x'-z_{i})$. Applying the Cauchy-Riemann conditions, the fact that the Green function satisfies,  $\nabla^{4}G(x,t ) =\frac{1}{\kappa^{2} } \delta^{(2)}(x)\delta(t) - \frac{1}{\kappa^{2} } \partial_{t}^{2} G(x,t),$ and the fact that the first set of $\delta$-functions are vanishing in the limit $r\ll R$, an integration by parts leaves the final result,
\begin{eqnarray}
	 \lim_{z_{i}\rightarrow z} \nabla^{2}_{(z_{1})} \nabla^{2}_{(z_{2})} \langle \tilde{\mathcal{O} }_{8}(\vec{x}) \tilde{\mathcal{O} }_{-8}(\vec{y})  :\varphi(\vec{z}_{1}) \varphi(\vec{z}_{2} ): \rangle &=& -m^{2} e^{- 2\pi m^{2} \kappa^{2} G_{reg}(x-y)} \Bigg[ \partial_{t} G(\vec{x}-\vec{z }, t_{x}-\tau) - \partial_{t} G(\vec{y}-\vec{z},t_{y}-\tau) \Bigg]^{2} \nonumber \\
 \end{eqnarray} 
Hence, one finds a familiar result. 
\begin{eqnarray}
	 \lim_{z_{i}\rightarrow z} \nabla^{2}_{(z_{1})} \nabla^{2}_{(z_{2})} \langle \tilde{\mathcal{O} }_{8}(\vec{x}) \tilde{\mathcal{O} }_{-8}(\vec{y})  :\varphi(\vec{z}_{1}) \varphi(\vec{z}_{2} ): \rangle =  - 16 m^{2} t^{2} e^{- 2\pi m^{2} \kappa^{2} G_{reg}(r)} \left( \frac{ e^{-R^{2}/4T} }{8\pi \ka T^{2} } \right)^{2} 
\end{eqnarray}

\subsubsection{Fusion into $(\nabla\varphi)^{4}$}

By a similar set of manipulations, this amounts to the computation,
\begin{eqnarray*}
	 && \lim _{z_{i}\rightarrow z} \partial_{i,(z_{1})} \partial^{i}_{(z_{2})}\partial_{j,(z_{3})} \partial^{j}_{(z_{4})} \langle \tilde{\mathcal{O} }_{8}(\vec{x}) \tilde{\mathcal{O} }_{-8}(\vec{y})  :\varphi(\vec{z}_{1}) \varphi(\vec{z}_{2} )\varphi(\vec{z}_{3}) \varphi(\vec{z}_{4}) : \rangle =e^{- 2\pi m^{2} \kappa^{2} G_{reg}(x-y)}   \\
	 && \lim_{z_{i}\rightarrow z} \partial_{i, (z_{1})} \partial^{i}_{(z_{2})} \partial_{j,(z_{3})} \partial^{j}_{(z_{4})} \Bigg[ \kappa^{2} \int d^{3}\vec{x}' \, \, \nabla^{4} \alpha(x') G(x'-z_{1}) -\alpha(z_{1} ) \Bigg]   \Bigg[ \kappa^{2} \int d^{3}\vec{x}'' \,\, \nabla^{4} \alpha(x'') G(x'' -z_{2} ) - \alpha(z_{2} ) \Bigg]  \\
	 &&\quad\times \Bigg[ \kappa^{2} \int d^{3}\vec{x}''' \,\, \nabla^{4} \alpha(x'') G(x'' -z_{3} ) - \alpha(z_{3} ) \Bigg] \Bigg[ \kappa^{2} \int d^{3}\vec{x}'''' \,\, \nabla^{4} \alpha(x'') G(x'' -z_{4} ) - \alpha(z_{4} ) \Bigg] 
\end{eqnarray*}
This is similar to the previous calculation, and one finds
 \begin{equation}
	\langle \tilde{\mathcal{O} }_{8}(\vec{x}) \tilde{\mathcal{O} }_{-8}(\vec{y})  :(\nabla\varphi)^{4}(\vec{z}): \rangle = \left( 4 \ka^{4} m^{2} \right)^{2} e^{- 2\pi m^{2} \kappa^{2} G_{reg}(r)} \left[ \frac{1}{(4\pi\ka)^{2}} 4r^{2} \frac{ f^{2}(s_{R})}{R^{4} } + \mathcal{O}(1/R^{3} ) + \mathcal{O}(1/R^{2}) \right]^{2} 
\end{equation}

\end{widetext}

\end{document}